\documentclass[fleqn,usenatbib]{mnras}

\usepackage{newtxtext,newtxmath}

\usepackage[T1]{fontenc}
\usepackage{ae,aecompl}

\usepackage{graphicx}	
\usepackage{amsmath}	
\usepackage{amssymb}	
\usepackage{tikz}
\usepackage[skins]{tcolorbox}

\title[RM-synthesis on SKA simulated galaxy clusters]{Rotation Measure synthesis applied on synthetic SKA images of galaxy clusters}

\author[Loi et al.]{
F. Loi$^{1,2}$\thanks{E-mail: francesca.loi@inaf.it},
M. Murgia$^{2}$,
F. Govoni$^{2}$,
V. Vacca$^{2} $,
A. Bonafede$^{1}$,
C. Ferrari$^{3}$,
\newauthor I. Prandoni$^{4}$,
L. Feretti$^{4}$, 
G. Giovannini$^{1}$,
and H. Li$^{5}$
\\
$^{1}$Dip. di Fisica e Astronomia, Universit\`a degli Studi Bologna, Viale Berti Pichat 6/2, I--40127 Bologna, Italy \\
$^{2}$INAF - Osservatorio Astronomico di Cagliari, Via della Scienza 5, I-09047 Selargius (CA), Italy \\
$^{3}$Observatoire de la Cote d'Azur, France\\
$^{4}$INAF - Istituto di Radioastronomia, Via Gobetti 101, I--40129 Bologna, Italy \\
$^{5}$Theoretical Astrophysics, Los Alamos National Laboratory, Los Alamos, NM\\
}

\date{Accepted XXX. Received YYY; in original form ZZZ}

\pubyear{2018}

\begin{document}
\label{firstpage}
\pagerange{\pageref{firstpage}--\pageref{lastpage}}
\maketitle

\begin{abstract}
Future observations with next generation radio telescopes will help us to understand the presence and the evolution of magnetic fields in galaxy clusters through the determination of the so-called Rotation Measure (RM).
In this work, we applied the $RM$-synthesis technique on synthetic SKA1-MID radio images of a pair of merging galaxy clusters, measured between 950 and 1750\,MHz with a resolution of 10$\arcsec$ and a thermal noise of 0.1$\muup$Jy/beam. The results of our $RM$-synthesis analysis are compared to the simulations' input parameters. We study two cases: one with radio haloes at the cluster centres, and another without. We found that the information obtained with the $RM$-synthesis is in general agreement with the input information. Some discrepancies are however present. We characterise them in this work, with the final goal of determining the potential impact of SKA1-MID on the study of cluster magnetic fields.
\end{abstract}



\section{Introduction}
\label{sec:intro}
The next generation of radio instruments are allowing us to investigate cosmic magnetism with unprecedented insight.
Indeed, our Universe is highly magnetised on all scales, from the smallest objects (stars and planets) to the largest (galaxies and galaxy clusters). Hints of the presence of magnetic fields are observed even on larger scales, namely in the filaments of the cosmic web \citep{govoni19,vacca18}.

Current models predict that magnetic fields in galaxy clusters could either originate from a primordial injection or from seeding from galactic outflows: the injected magnetic field would then be amplified and spread all over the intracluster-medium (ICM) via turbulence and shock waves, up to a strength of few $\muup$G on Mpc-scales \citep{donnert}. In order to distinguish between different scenarios, it is important to study the magnetic fields in galaxy clusters and filaments both through numerical simulations and observations.\\
The effect of Faraday rotation on background and cluster radio sources \citep[e.g.][]{clarke, govoni04} is very useful to determine the intracluster magnetic fields properties. The Faraday effect manifests as the rotation of the polarization plane of a linearly polarized signal passing through a magneto-ionic medium and depends as a function of the square of the wavelength, $\lambda^2$, and the Faraday depth $\phi(l)$ as follows:
\begin{equation}
    \Delta \Psi = \Psi-\Psi_0=\phi(l) \cdot \lambda^2.
    \label{eq:far}
\end{equation}
The Faraday depth is the integral of the line-of-sight parallel component of the intervening magnetic field multiplied by the thermal plasma density:
\begin{equation}
\phi(l)=\int_0^l B_{||} \cdot n_e dl,
\label{eq:rm}
\end{equation}
where the integral is performed over the path crossed by the signal. In the case of a radio source whose signal passes through the magneto-ionic medium of a galaxy cluster and its observed polarized emission follows the $\lambda^2$-dependence of Eq. \ref{eq:far}, the associated Faraday depth is known as cluster's Rotation Measure (RM) and the information it encodes is linked to the properties of that cluster, particularly its intracluster magnetic field.
Evaluating the Faraday depth is useful to investigate the magnetic field on small scales ($\sim$\,kpc), but its exploitation is limited by the current sensitivity of radio telescopes, which does not allow us to detect more than a few of polarized radio sources per cluster and only in a few clusters. Indeed, the densest $RM$ Grid at 1.4\,GHz available today has an average density of 1\,RM source per square degree \citep{taylor}, and it is based on the NRAO VLA Sky Survey \citep[NVSS,][]{condon98}.  \\
Some galaxy clusters show a central diffuse radio emission which is not associated with specific optical counterparts or discrete radio sources. The observed properties of these so-called radio haloes \citep[see the reviews of][]{van19,feretti12,ferrari08} reveal their synchrotron nature, and prove the existence of a non-thermal component in the ICM, made of relativistic particles and large scale magnetic fields.
Radio haloes are found at the centre of merging clusters, and are typically ${\rm \sim1\,Mpc}$ in size. They are characterised by a low surface brightness (${\rm S_{\nu} \sim 0.1-1\,\muup Jy/arcsec^{2}}$ at 1.4\,GHz), with a steep power-law spectrum (${\rm S_{\nu}=S_0 ~ ( {\nu}/{\nu_0})^{-\alpha}}$, ${\rm \alpha \gtrsim1}$).
Although radio haloes are synchrotron sources, they are generally observed to be unpolarized, though polarized radio halo emission as been detected (up to $\sim$20\% at 1.4\,GHz), in galaxy clusters A2255 \citep{a2255}, MACS J0717.5+3745 \citep{macs}, and A523 \citep{a523}. The absence of polarized emission in radio haloes could be attributed to two types of depolarization effects: internal and external depolarization. External depolarization is caused by instrumental effects. It occurs when the emitting magnetic field is tangled on a scale smaller than the beam area of the telescope and it can be observed also in the case of discrete radio sources. In addition, Faraday rotation can take place within a frequency channel and the larger the channel, the larger the rotation. The result is an incoherent sum of polarized signals that, averaged together, effectively suppress the measured degree of polarization. Internal depolarization takes place whenever a radio emitting source is mixed with a magneto-ionic medium: every polarized signal emitted in a different position along a given line-of-sight will experience a different Faraday depth, corresponding to the crossed portion of the plasma; thus, all the signals coming from that line-of-sight will be rotated by different angles and when they reach the telescope they sum up incoherently in each frequency channel. As a result the degree of polarization is strongly reduced at long wavelengths with respect to the intrinsic value that would be observed at $\lambda=0$.\\ 
To mitigate the impact of the problems mentioned above, it is common to use the $RM$-synthesis technique \citep{burn,brent}. This consists of building a spectrum of the polarization as a function of the Faraday depth, known as the Faraday dispersion function  $F(\phi)$. This spectrum is built for each given line-of-sight using the $Q$ and $U$ Stokes parameters measured by an instrument. Assuming a value of
Faraday depth $\phi(l)^{\ast}$ in a given interval, $Q$ and $U$ are de-rotated in each frequency channel according to this value. The polarized intensity associated with this Faraday depth is then computed as:
\begin{equation}
    ||P||=\sqrt{Q^2+U^2},
    \label{eq:p}
\end{equation} 
being $P=Q+iU$, and $Q$ and $U$ real quantities.
By repeating this procedure for different values of Faraday depths inside a given interval, it is possible to obtain an approximate reconstruction of the Faraday dispersion function. More formally, it can be shown that a Fourier relation exists between the polarization distribution in Faraday space and the observed polarization in the frequency domain. If this technique is applied to a background radio galaxy, in absence of any Faraday screen except the cluster ICM itself, the result is a Faraday dispersion function with a single peak at the Faraday depth which corresponds to the cluster $RM$. In the case of radio haloes, or emitting sources mixed with rotating plasma, or multiple sources along the line-of-sight, there will be several polarized peaks emitted at different position along the path towards the telescope which will experience different Faraday depths. The Faraday dispersion function
can thus feature multiple components, and determining which is associated
with the cluster we are interested in can be a non-trivial task.

In this work, the $RM$-synthesis is applied to synthetic radio images of galaxy clusters. These are the results of a cosmological magneto-hydro-dynamical (MHD) simulation of the ICM properties of a pair of merging galaxy clusters in combination with the {\sc faraday} software package \citep{murgia04}. In previous works, this tool has been shown to be capable of reproducing the radio emission associated to a discrete radio source population \citep{loi19} and to cluster radio haloes \citep{govoni13,xu}. Here, we focus on investigating the potential benefits of the $RM$-synthesis technique applied to synthetic SKA1-MID data, aiming to determine the cluster $RM$. Throughout the paper, the following nomenclature is adopted: {\it simulated} images are images at the maximum resolution offered by the MHD simulations, i.e. $\sim$10.7\,kpc (which corresponds to $\sim5\arcsec$ at the clusters redshift z=0.115). The thermal noise in our simulated images is zero; {\it synthetic} images are the simulated images convolved with a Gaussian function having FWHM=10$\arcsec$ after adding a thermal noise of 0.1$\muup$Jy/beam. 
{\it Intrinsic} parameters refer to those physical observables that can be computed from the MHD cubes, such as the cluster $RM$.
The simulated and synthetic images are produced considering the frequency band 950$-$1760\,MHz (i.e. the band 2 of the SKA1-MID). We treat two separate cases: clusters with radio haloes, and clusters without radio haloes. The results of the $RM$-synthesis applied to the data are compared with the intrinsic cluster $RM$.

The paper is organised as follows: 
in Section 2, we describe the assumptions made to produce the simulated full-Stokes images of the two merging clusters; in Section 3, we show the simulated and synthetic total intensity radio images. Section 4 describes the application of the $RM$-synthesis technique on simulated and synthetic data, along with how the results will be treated. Section 5 reports the results of the $RM$-synthesis on simulated and synthetic images. Finally, Section 6 is devoted to the analysis and Section 7 to the conclusions. In the Appendix, a simple estimate of the intracluster magnetic field strength is given.\\
Throughout the paper, a ${\rm\Lambda}$CDM cosmology is adopted with ${\rm H_0=71\,km \cdot s^{-1} Mpc^{-1}}$, ${\rm\Omega_m=0.27}$, and ${\rm\Omega_{\Lambda}=0.73}$. The galaxy clusters of this work are simulated at redshift z=0.115 where 1$\arcsec$=2.06\,kpc.

\begin{figure*}
    \centering
    \includegraphics[width=0.33\textwidth]{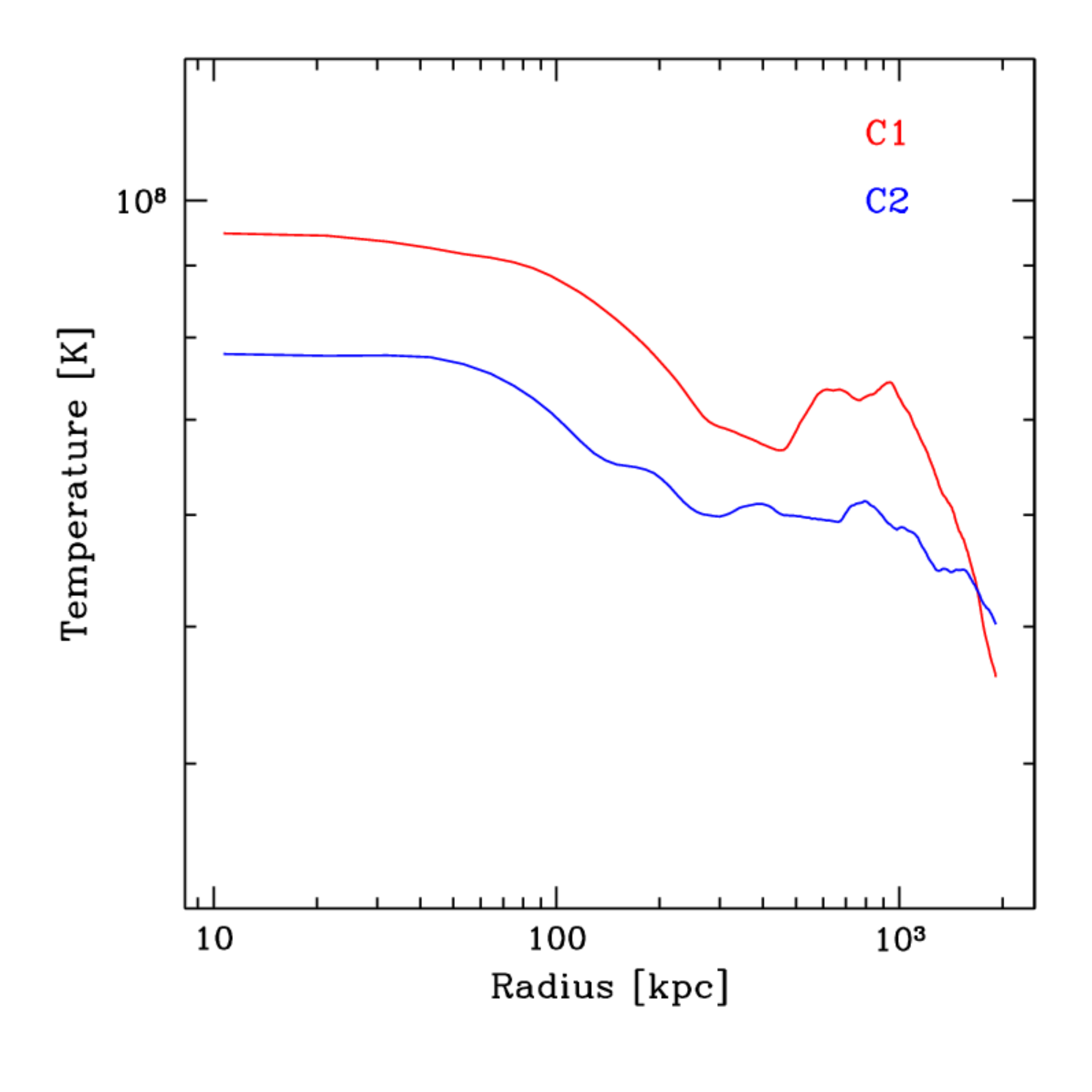}
    \includegraphics[width=0.33\textwidth]{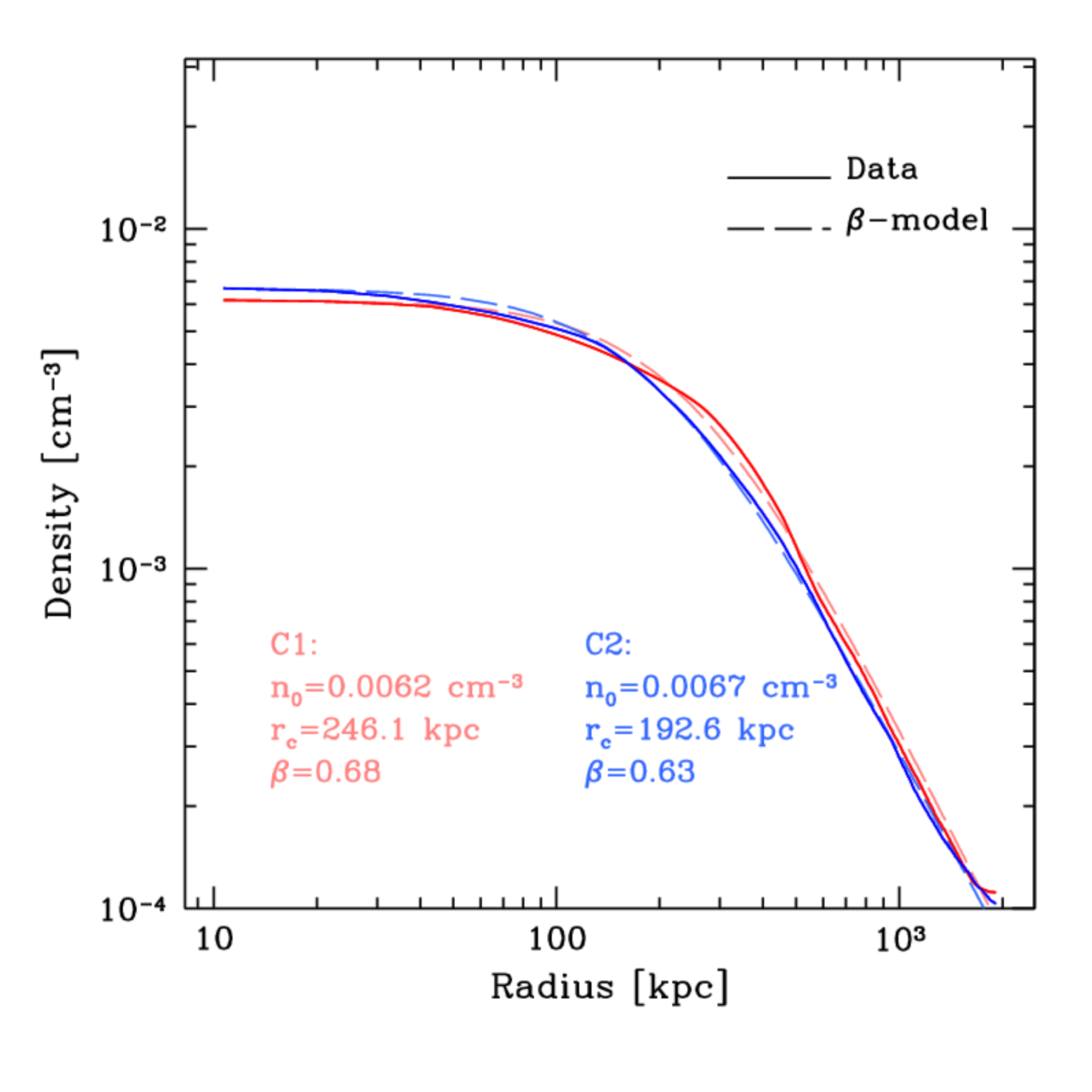}
    \includegraphics[width=0.32\textwidth]{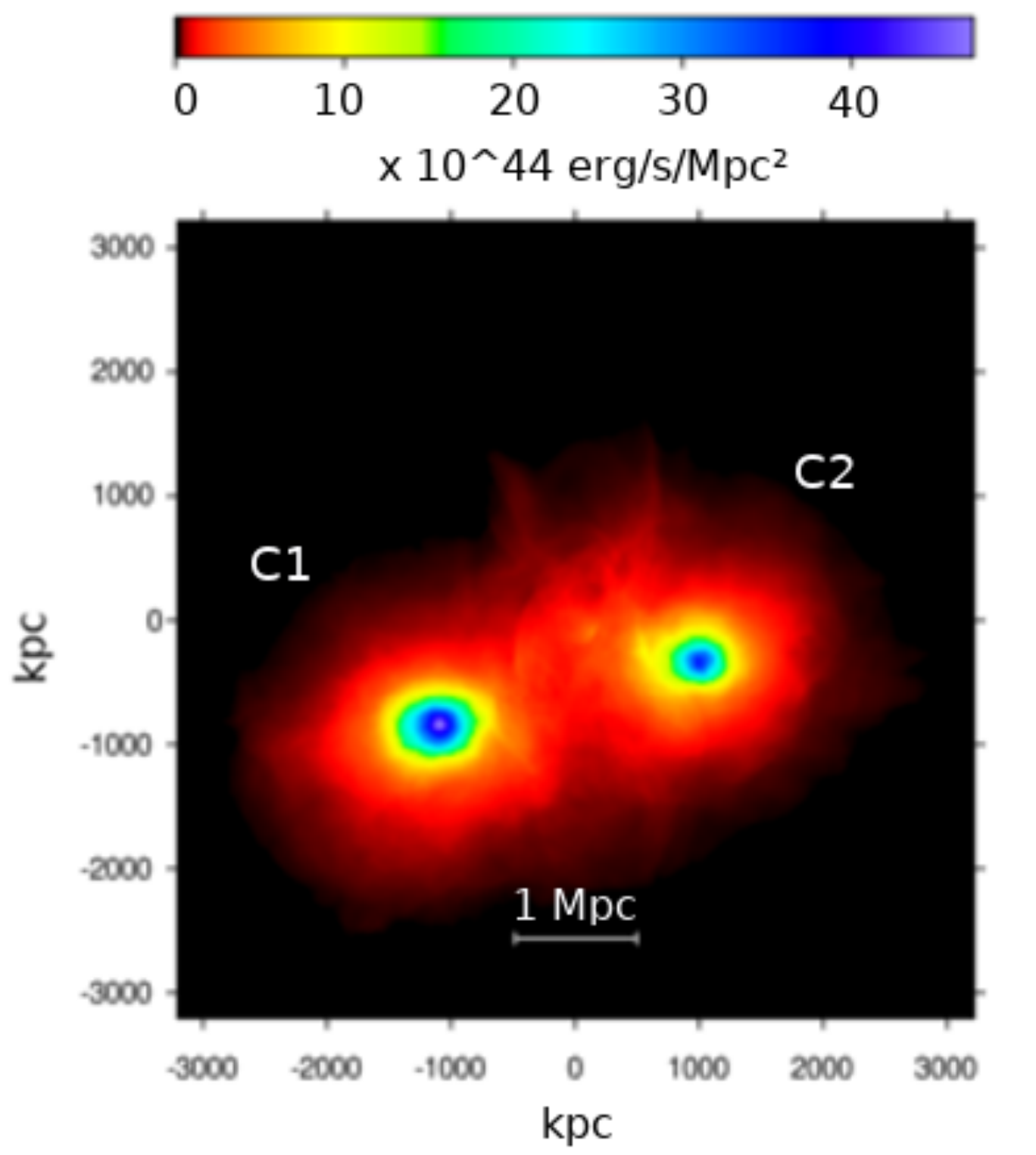}
       \caption{From left to right: temperature, and density profiles for the C1 (red) and C2 (blue) clusters, and X-ray surface brightness of the two clusters in the 0.1-2.4\,keV band. For the density profile the best-fit equations of a $\beta$-model are plotted as dashed lines with parameters reported in the bottom of the plot.}
    \label{fig:tdb}
\end{figure*} 
\section{Characteristics of the simulations}
In this Section, we describe the models and simulations adopted to produce full-Stokes spectro-polarimetric cubes of the merging system. Cluster ICM properties (i.e. thermal density, temperature, and magnetic field) are derived from cosmological MHD simulation, while the total and polarized radio emission are simulated using the previous properties.

\subsection{MHD simulations}
The cosmological MHD simulation presented here is obtained with the {\sc enzo} code \citep{collins} with adaptive mesh refinement (AMR) by the group of Hui Li at the Los Alamos National Laboratories, USA. It is a small part of a bigger cosmological simulation realised in a volume of (256 ${\cdot\, h^{-1}\,\rm Mpc^{-3}}$) which encloses several forming structures. The evolution of the dark matter, baryonic matter, and magnetic fields are built in the simulation. 
An adiabatic equation of state with a specific heat ratio ${\rm\Gamma=5/3}$ is used, while heating and cooling physics or chemical reactions are not included. The simulation runs from redshift z=30 to z=0. 
For this work, we focus on a pair of galaxy clusters with similar masses that undergoes a merging process resulting in a single system at z=0.073, with a total mass of ${\rm M=1.9\times10^{15} M_{\odot}}$. The process lasts for about 2\,Gyrs.
The magnetic fields are injected by AGNs at z=2-3 \citep{xu} and then amplified and spread over Mpc-scales during the late stages of the merger.\\
For the purposes of this work, a single snapshot of the MHD simulation at z=0.115 is used. At this particular instant the clusters centres are separated by about 2\,Mpc.
The simulation consists of a set of 3-dimensional cubes of ${\rm \sim (6.4\,Mpc)^3}$ with a cell size of ${\rm \sim 10.7\,kpc}$ containing the ICM physical parameters: temperature, thermal plasma density, and intracluster magnetic fields.
The physical configuration is very similar to that observed for the pair of galaxy clusters A399-A401 where both the systems host a diffuse radio halo \citep{murgia2010}.
\begin{table}
	\centering
	\caption{Parameters of the MHD simulation: from left to right, the field-of-view (FOV), pixel size, mass, and $r_{200}$ of cluster C1 (first row) and C2 (second row) are listed. The $\beta$-model parameters, namely the central density $n_0$, the core radius $r_c$, and $\beta$, are also listed in this table.}
	\label{tab:sim}
	\begin{tabular}{l l l l l l l} 
		\hline
		FOV & pix.size & mass & r$_{200}$ & n$_0$ & r$_c$ & $\beta$ \\	
		$\rm[Mpc]$ & $\rm[kpc]$ & $\rm[10^{14}M_{\odot}]$ & $\rm[Mpc]$ & $\rm [cm^{-3}]$ & $\rm[kpc]$ &  \\
		\hline
		6.4$^2$ & 10.7 &  4.7 & 1.2 & 0.0062 & 246.1 & 0.69\\
		& & 3.8 & 1.1 & 0.0067 & 192.6&  0.63\\
		 \hline 
	\end{tabular}
\end{table}

The characteristics of the simulation are summarised in Table \ref{tab:sim}, where we also specified the $r_{200}$ radius, defined as the distance within which the cluster density is 200 times the critical density of the Universe.\\

\subsubsection{Temperature, density, and X-ray emission}
Fig. \ref{fig:tdb} shows the temperature (left), and density (middle) profiles as a function of the distance from the cluster centres, computed from the 3 dimensional MHD cubes, for the C1 (red) and the C2 (blue) clusters. 
The simulated ICM reaches temperature as high as $\sim 8-9 \cdot10^7$\,K in the cluster centre, decreasing towards the outskirts of the cluster both for C1 and C2, even if a clear enhancement can be observed in C1 at $\sim$700\,kpc from the cluster centre.
At the cluster centres, the density assumes values of the order of $\sim 6\cdot10^{-3} \rm cm^{-3}$, decreasing going to large distances. 
Even if this is a merging system, the density profiles are well described by a $\beta$-model \citep{cavaliere}:
\begin{equation}
    n(r)=n_0 \cdot \left ( 1+ \left (\frac{r}{r_c} \right )^2 \right )^{-3\beta/2},
    \label{eq:bmodel}
\end{equation}
with best-fit parameters for the central density $n_0=0.0062 \,\rm cm^{-3}$, core radius $r_c=246.1\,$kpc, and $\beta=0.69$ for the C1 cluster and $n_0=0.0067 \,\rm cm^{-3}$, $r_c=192.6\,$kpc, and $\beta=0.63$ for the C2 cluster.
These parameters are reported in Table \ref{tab:sim}. The best-fit equations are plotted as dashed lines.\\
From the MHD cubes it is possible to evaluate the X-ray surface brightness of the clusters in the energy band 0.1-2.4\,keV. This is shown on the right of Fig. \ref{fig:tdb}.

\begin{figure*}\centering
   \includegraphics[width=0.34\textwidth]{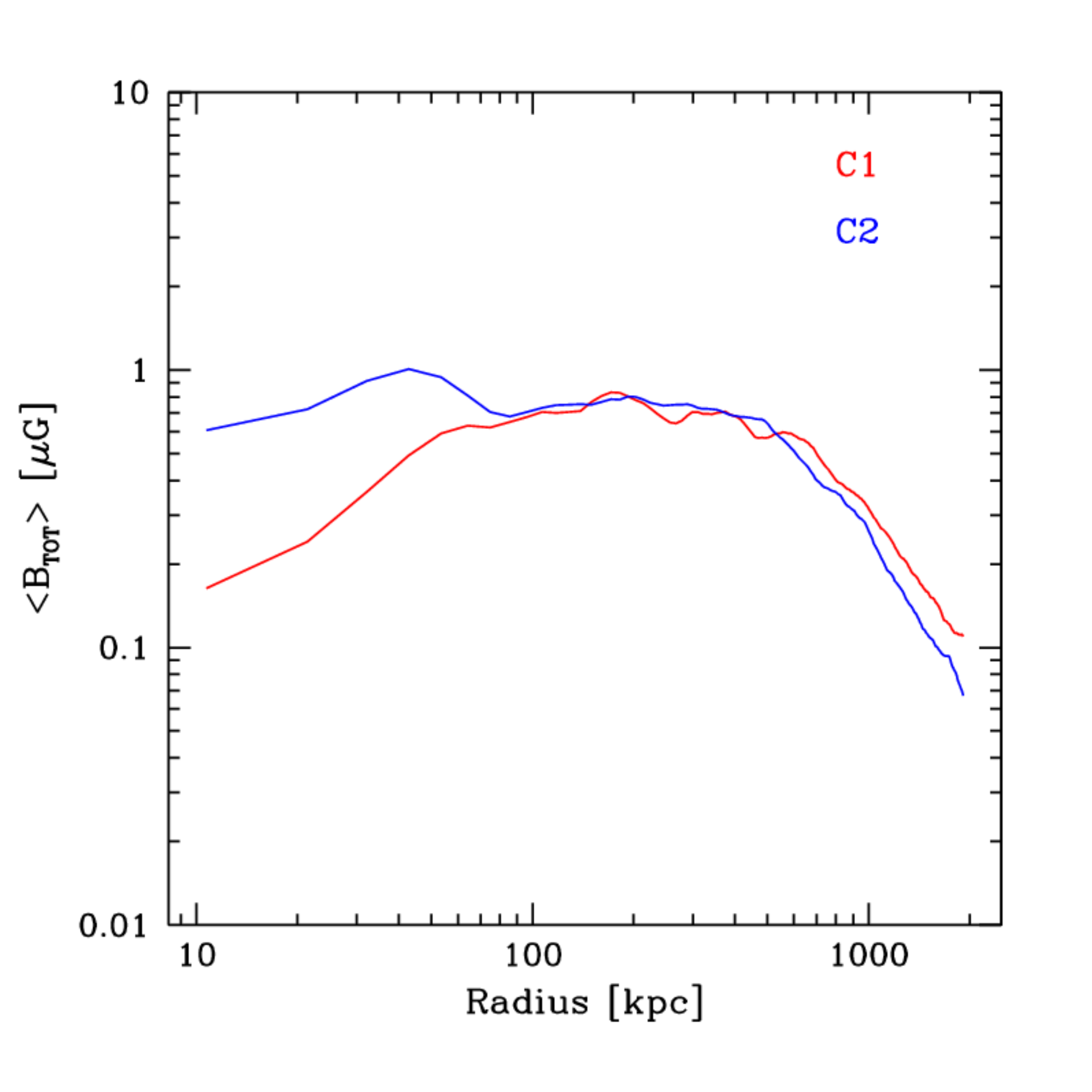}
   \includegraphics[width=0.34\textwidth]{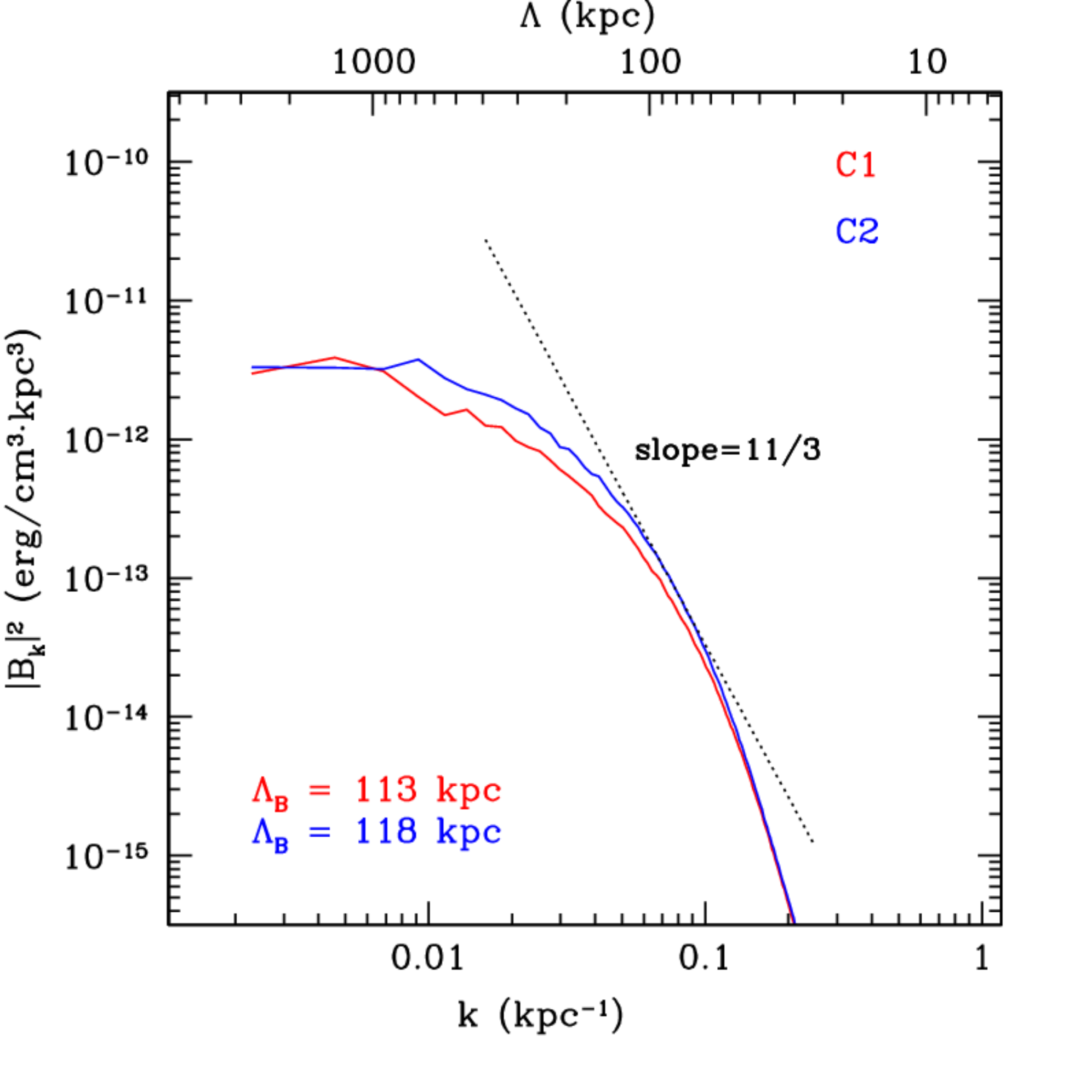}
    \caption{Left: average of the total magnetic field 3D profiles for the C1 (red), and the C2 cluster (blue). Right: magnetic field power spectra for the C1 (red), and the C2 cluster (blue). The autocorrelation length values are shown in the bottom left corner. A dotted black lines shows a power spectrum with spectral index equal to 11/3.}
    \label{fig:ps}
\end{figure*}
\begin{figure*}
    \centering
    \includegraphics[width=0.33\textwidth]{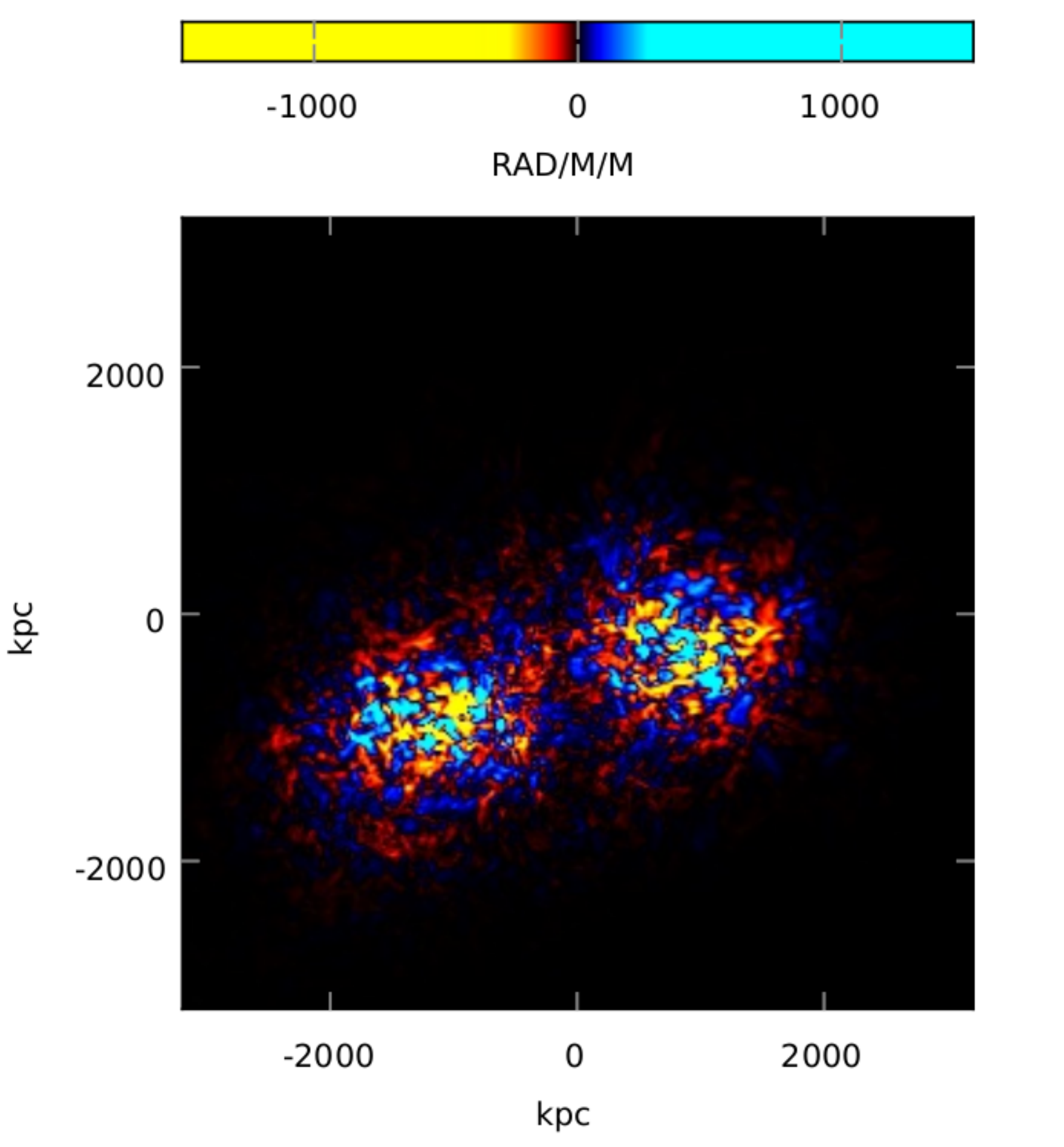}
    \includegraphics[width=0.32\textwidth]{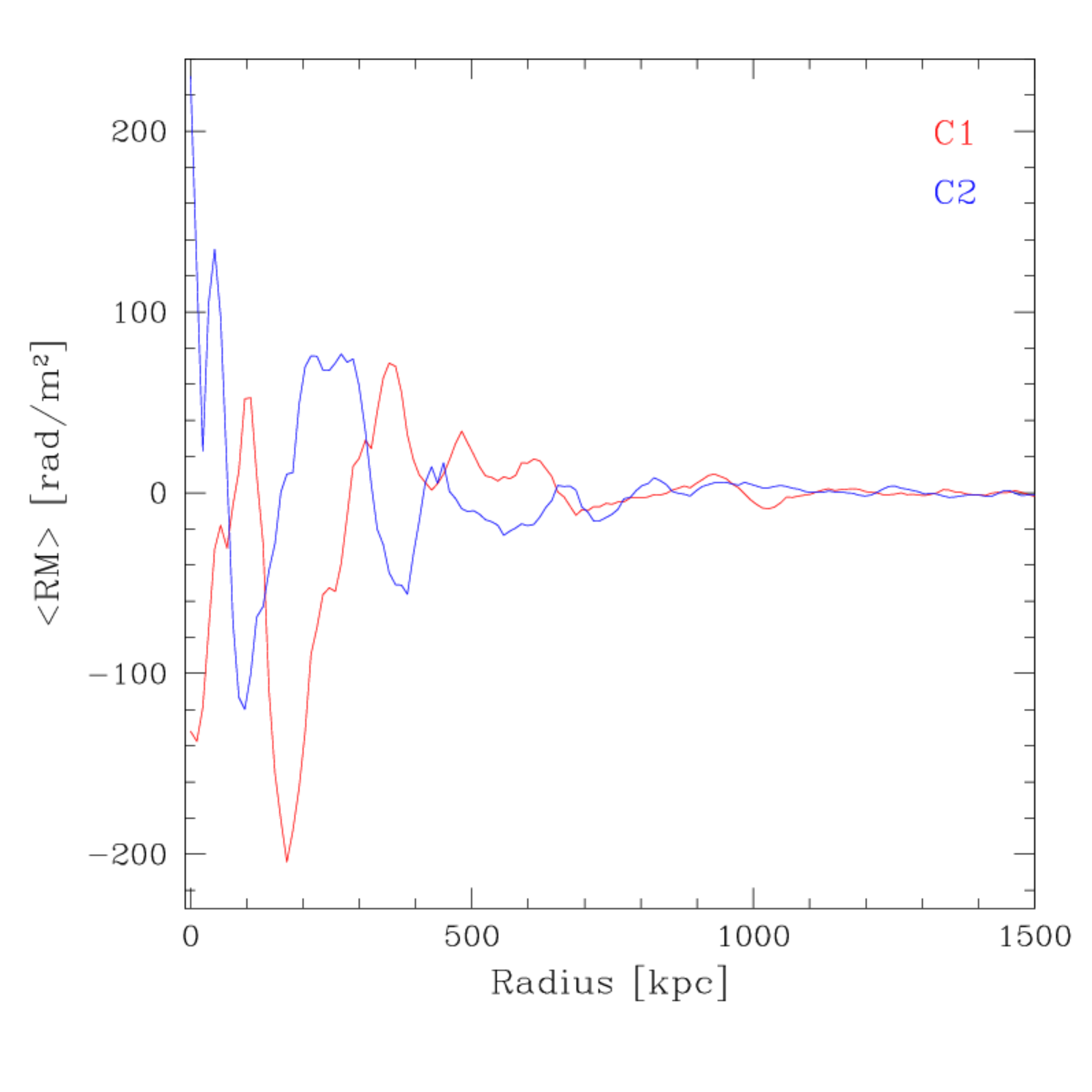}
    \includegraphics[width=0.32\textwidth]{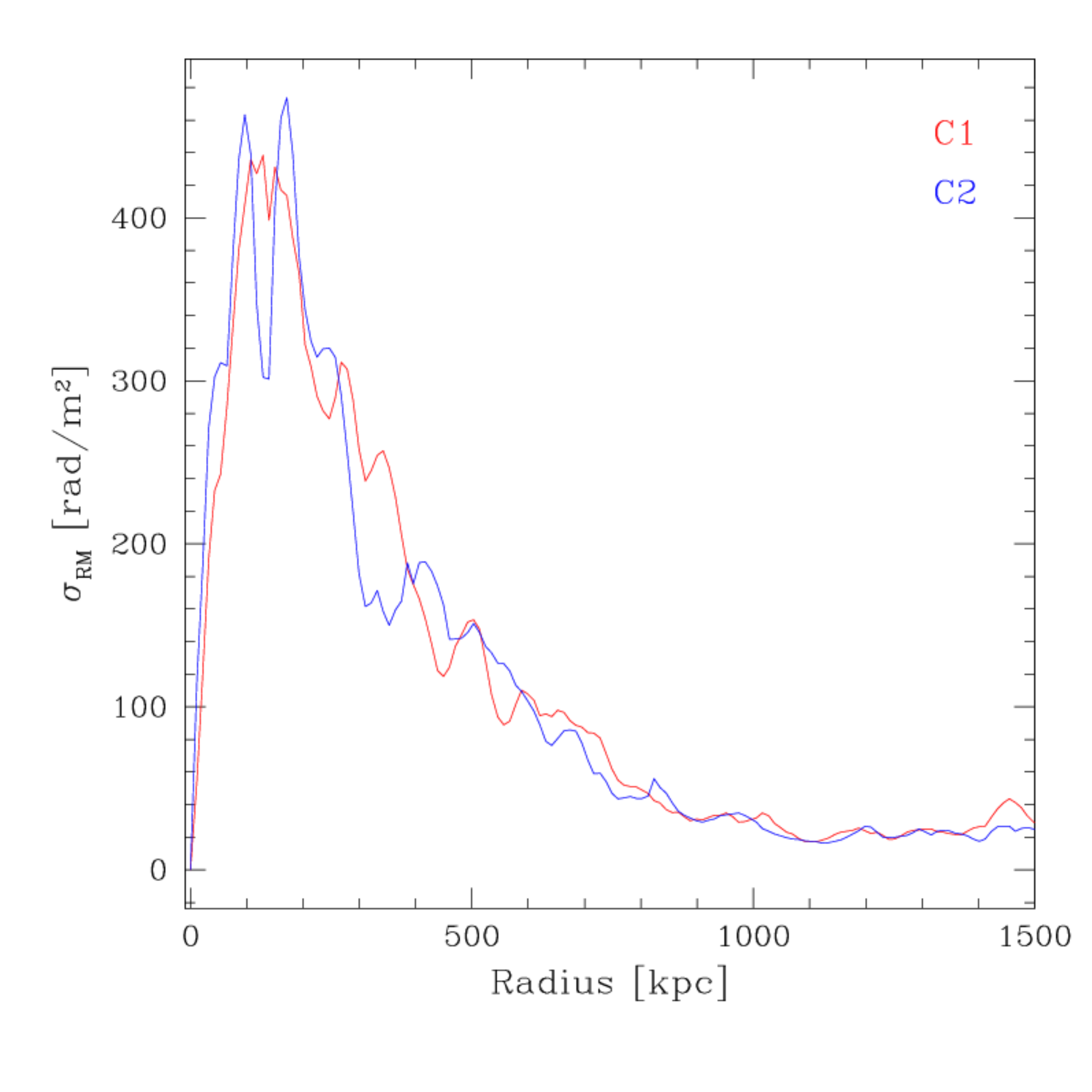}
    \caption{From left to right: intrinsic $RM$ image computed from the MHD cubes, azimuthally averaged profiles in circular annuli of the mean $RM$ and of the $\sigma_{RM}$ as a function of the distance from the cluster centres, in red for the C1 cluster and in blue for the C2 cluster.}
    \label{fig:rm}
\end{figure*}

\subsubsection{Magnetic field and $RM$ properties}
Fig. \ref{fig:ps} shows on the left the average of the magnetic fields profile of the C1 (red) and the C2 (blue) clusters. Their intensities range from 0.1 and 1\,$\muup$G.
The C1 cluster magnetic field shows an enhancement of its intensity from $\sim$\,0.2 up to $\sim$0.7\,$\muup$G in the range between 10\,kpc 100\,kpc and then a decrease at distances larger than $\sim$500\,kpc.
The C2 cluster has a flat behaviour with an intensity of $\sim$0.6-0.7\,$\muup$G up to 500\,kpc and then a decrease similar to the one observed in the C1 cluster.\\
Fig. \ref{fig:ps} also shows, on the right, the magnetic field power spectra of the two clusters, computed as the Fourier Transform of the 3D magnetic field autocorrelation function in the wave number domain $k=2\pi/\Lambda$, being $\Lambda$ the scale in kpc. As before, the power spectra of the C1 and C2 clusters are respectively traced in red and blue. For both the clusters, the magnetic field autocorrelation lengths have been evaluated from \citep{enss03}:
\begin{equation}
    \Lambda_B=\frac{3\pi}{2}\frac{\int_0^{\infty} | B_k |^2 \, k \,dk}{\int_0^{\infty} | B_k |^2\, k^2\, dk},
\end{equation}
and their values are reported in the bottom left corner of the plot. For the sake of comparison, a dotted black line traces a Kolmogorov-like power spectrum (slope equal to 11/3).

The cluster $RM$ can be computed from Eq. \ref{eq:rm} with the integral performed across the entire MHD box size. The resulting image is shown in Fig. \ref{fig:rm} (left panel) together with the profile of the mean $RM$ (middle panel), and its standard deviation $\sigma_{RM}$ (right panel). 
This image suggests a turbulent magnetic field structure. From the profiles, we can observe large oscillations of the $RM$ mean and high values of $\sigma$ near to the centre, both decreasing as a function of distance. However, at the centre the profiles go to zero due to the fact that the autocorrelation scale of the $RM$ is larger than the size of the first annulus.

\subsection{{\sc faraday} simulations}
The {\sc faraday} software package \citep{murgia04} has been specifically designed for intracluster magnetic fields investigations. It can reproduce spectral cubes of the total and polarized intensity of simulated radio haloes \citep{murgia04,govoni13}, as well as the radio emission of a population of discrete radio sources \citep[see][for more details]{loi19}.
Here, the discrete radio sources populating the field-of-view are simulated according to \citet{loi19}. The radio haloes are simulated under two assumptions: 1) equipartition between magnetic field and relativistic electron energy density voxel by voxel, consistently with the work of \citet{govoni13}, 2) coupling between the relativistic particles responsible for the radio halo emission and the thermal particles. In this latter scenario the energy density of the relativistic particles constitutes the 0.3\% of the thermal energy density. This factor has been chosen in order to obtain radio power at 1.4\,GHz similar to that of radio haloes in equipartition condition, and it is consistent with the upper limit set from $\gamma$-ray observations \citep[ratio between the relativistic and thermal particle energy density less than 10\%, see][]{brunetti17}.\\
To complete these radio simulations and make them suitable in the case of a pair of galaxy clusters, it is necessary to include a cluster population of radio sources. 

\subsubsection{Cluster discrete radio sources}
The cluster population is added to the simulation as follows:
\begin{list}{}{}
 \item[1)] two families of sources are considered: active galactic nuclei (AGN) and star forming galaxies (SFGs);
 \item[2)] values for ${\rm r_{200}}$ and for the cluster volume are set;
 \item[3)] it is assumed that the spatial distribution of the radio sources follows a Navarro-Frank-White profile \citep{lin} and that the luminosity and density of the cluster sources are described by the radio luminosity functions of the same work;
 \item[4)] the Navarro-Frank-White distribution is normalized so as to yield the same number of sources per Mpc$^3$ provided by the radio luminosity function; 
 \item[5)] the total number of cluster sources N is computed from the integration of the spatial distribution over the cubical box;
 \item[6)] the number of AGN or SFGs is computed by multiplying the total number N by the ratio between the radio luminosity function of one specific type (AGN or SFG) with respect to the total radio luminosity function;
 \item[7)] by using a Monte Carlo approach the position (x,y,z), and the luminosity of each source are extracted from the corresponding cumulative distribution functions.
\end{list}
The size, the morphology, and the spectro-polarimetric properties of the sources are assigned as described in \citet{loi19} (see that paper for more details). 
It is worth recalling that radio galaxies moving across the ICM typically show a distorted morphology caused by the jets bending. Such tailed radio galaxies are classified as narrow-angle tail, where the jets of a Fanaroff-Riley (FR) type I \citep{fr} are evidently bent, and wide-angle tail, less bent than the previous one. When the jets of a narrow-angle tail galaxies point toward the same direction the radio source is called head-tail galaxy. The simulations presented in this paper show cluster radio galaxies with such peculiar morphology.

\subsubsection{Polarized emission}
\label{subs:pol}
The polarized emission of radio sources, and the resulting images, are obtained through the following steps for each line-of-sight:
\begin{enumerate}
    \item from the intrinsic polarized intensity $||p_{\nu}||$ and angle $\psi_0$ of a radio source at a given depth $l$, the $Q_{\nu}(l)$ and $U_{\nu}(l)$ Stokes parameters are computed taking into account the corresponding Faraday depth (see Eq. \ref{eq:far}), according to:
    \begin{eqnarray}
    Q_{\nu}(l) & = &  \frac{||p_{\nu}||}{\sqrt{\tan^2{2\Psi}+1}} \nonumber \\
    U_{\nu}(l) & = &  \frac{||p_{\nu}||\tan{2\Psi}}{\sqrt{\tan^2{2\Psi}+1}},
    \end{eqnarray}
    at every $\nu$ of the simulated bandwidth;
    \item the $Q_{\nu}(l)$ and $U_{\nu}(l)$ contributions are then summed up along the line-of-sight:
    \begin{eqnarray}
        Q_{\nu}&=&\int_0^L Q_{\nu}(l) dl \\ \nonumber
        U_{\nu}&=&\int_0^L U_{\nu}(l) dl
    \end{eqnarray}
\end{enumerate}{}
By repeating this procedure for each line-of-sight it is possible to obtain cubes of the $Q_{\nu}$ and $U_{\nu}$ Stokes parameters. The polarized intensity images are obtained by averaging the $Q$ and $U$ cubes over the whole frequency band and then computing the polarized intensity using Eq. \ref{eq:p}.  
The images are then produced taking into account the depolarization effects mentioned in Section \ref{sec:intro}. It is worth noticing that they correspond to the result of the $RM$-synthesis considering the polarized signal at Faraday depth equal to zero. Such images will be compared in Section \ref{sect:rmres} with the linearly polarized images retrieved after the application of the $RM$-synthesis. 

\section{Total intensity radio images}
Starting from the MHD cubes, and making use of the source modelling implemented in {\sc faraday}, it is possible to reproduce the radio emission of the sources populating the field-of-view of the simulated cubes considering a given observational set up. In the following Section, we show simulated and synthetic total intensity images. The firsts are at the full resolution allowed by the simulation, $\sim 5\arcsec$/pix, and noise free, while the second are convolved with a beam resolution of 10$\arcsec$ and a thermal noise of 0.1\,$\muup$Jy/beam is added.
\begin{figure*}
    \centering
    \includegraphics[width=1\textwidth]{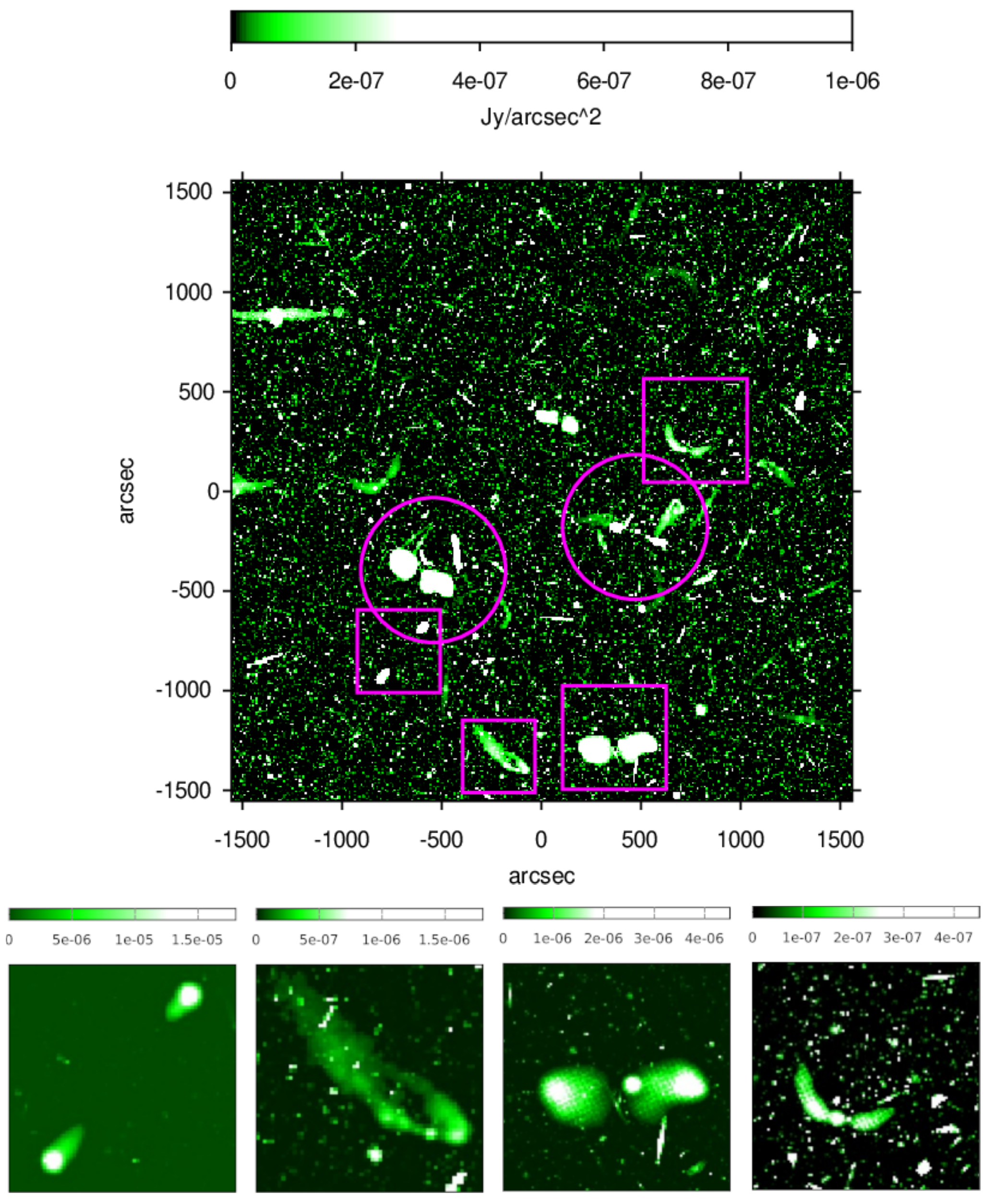}
    \caption{1.4\,GHz total intensity image of the two simulated galaxy clusters with a cell size of $\sim$5$\arcsec$. The magenta circles are centred on the cluster centres and have a radius 1.5\,Mpc. The magenta boxes indicate four different cluster radio sources that are shown in the bottom zoomed panels (see the text for more details). The surface brightness in the images goes from 0 up to $\sim$0.5\,mJy/arcsec$^2$ and it is obtained by spectral averaging all the frequency channels.}
    \label{fig:i_full}
\end{figure*}
\begin{figure*}
    \includegraphics[width=0.66\textwidth]{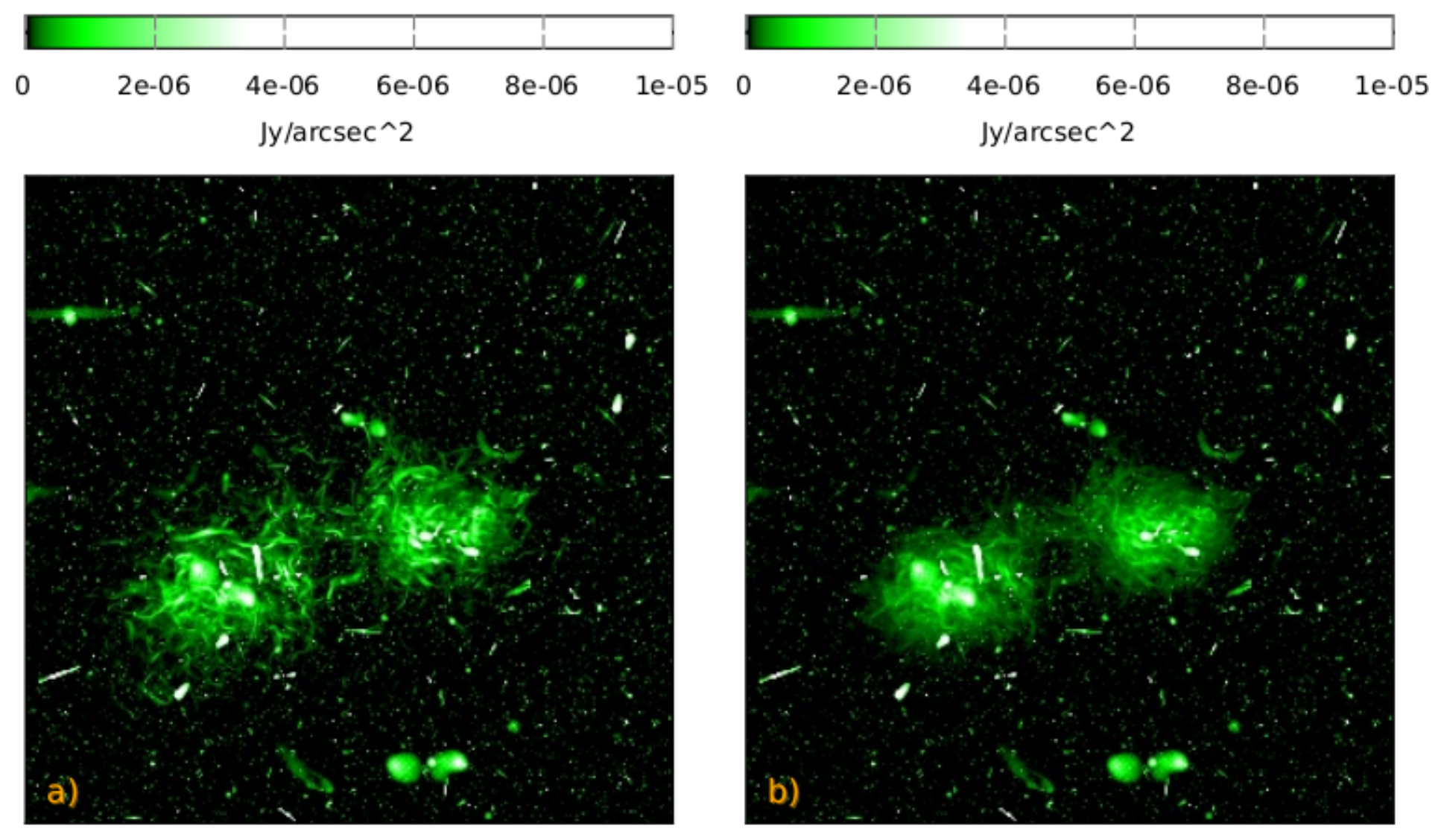}
    \includegraphics[width=0.33\textwidth]{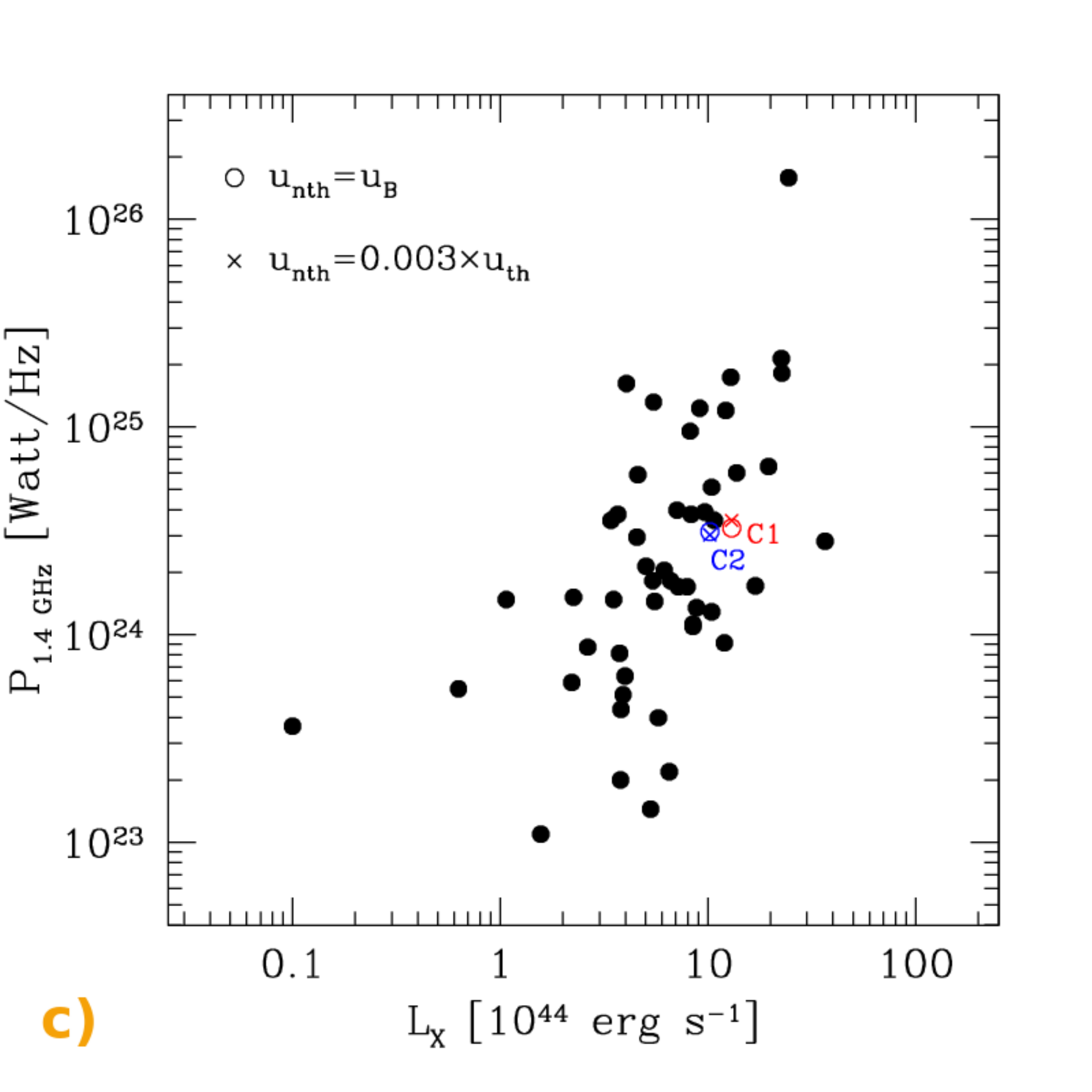}
    \caption{1.4\,GHz total intensity image of the two simulated galaxy clusters with a cell size of $\sim$5$\arcsec$ where (a) radio haloes are simulated according to the equipartition condition, and (b) the relativistic particles energy density is 0.3\% of the thermal plasma energy density. c) panel shows the 1.4\,GHz power versus X-ray luminosity of several haloes reported in the literature (black) and those simulated here in red for the C1 cluster and in blue for the C2 one. Open circles and crosses refer to the two assumptions made to simulate the radio haloes. The surface brightness in the images goes from 0 up to $\sim$0.5\,mJy/arcsec$^2$.}
    \label{fig:i_h}
\end{figure*}

\subsection{Simulated images}
Fig. \ref{fig:i_full} (top panel) shows the total intensity emission produced from a data cube spanning the frequency band 950-1760\,MHz with a spectral resolution of 1\,MHz. These simulated images are at the maximum resolution offered by the MHD simulations, i.e. $\sim$10.7\,kpc which corresponds to $\sim5\arcsec$ at the clusters redshift z=0.115, where the thermal noise is zero.\\
The dynamical range is very high going from the sub-$\muup$Jy/arcsec$^2$ brightness of the most distant sources up to the mJy/arcsec$^2$ level of the cluster sources. The maximum value of the surface brightness is $\sim$0.5\,mJy/arcsec$^2$, but it is not represented in the colour bar which would be otherwise highly saturated.
Two magenta circles of radius 1.5\,Mpc are centred on the cluster centres.
The magenta boxes in the images identify the four different cluster radio sources shown in bottom panels: from left to right, a FRII, an head-tail, a cluster wide-angle tail, and a FRII radio galaxy.

Fig. \ref{fig:i_h} shows the total intensity image again, this time with the addition of the two clusters radio haloes. In panel a), we assume the equipartition condition between the energy density of relativistic particles and the intracluster magnetic field. In panel b), we assume that the relativistic particles' energy density is 0.003 times the thermal plasma energy density, a factor chosen to obtain radio haloes with same luminosity at 1.4\,GHz. In these images, the colour bar limits were chosen to improve the contrast, but the actual maximum pixel values can reach $\sim$0.5\,mJy/arcsec$^2$. The surface brightness of radio halos present high values at the centre, between 0.3 and 0.7\,$\muup$Jy/arcsec$^2$, decreasing outwards. 
The filamentary morphology of the simulated radio haloes reflects the intracluster magnetic field structure, especially when equipartition is assumed. The radio haloes generated coupling between the thermal and non-thermal particles energy density create a smoother morphology. This is because the emissivity of radio haloes in equipartition, assuming $\alpha=1$, is proportional to $B^4$, while it is proportional to $B^2$ for radio haloes coupled with thermal particles. This results in emission of radio haloes under equipartition tracing the turbulent nature of the magnetic field more dramatically.\\
The properties of the simulated radio haloes are similar to those observed in terms of radio power at 1.4\,GHz and X-ray luminosity. This is shown in the right panel of Fig. \ref{fig:i_h}, where the two open circles and crosses correspond to the radio haloes in equipartition and coupled with the thermal plasma respectively, in red for the C1 and in blue for C2 cluster haloes, while the black points represents the results obtained for several observed radio haloes \citep[see the compilation of][]{feretti12,fede12,van12,giovannini,martinez,shakouri16,parekh,loi}. It is worth noting that, even if the integrated values of the 1.4\,GHz radio power are the same under both assumptions, the surface brightness of the radio haloes simulated coupling between relativistic and thermal particles show a rapid decrease going outwards, whereas under equipartition bright filaments are still visible at distances larger than 1.5\,Mpc from the cluster centres.
\begin{figure*}
    \centering
    \includegraphics[width=0.9\textwidth]{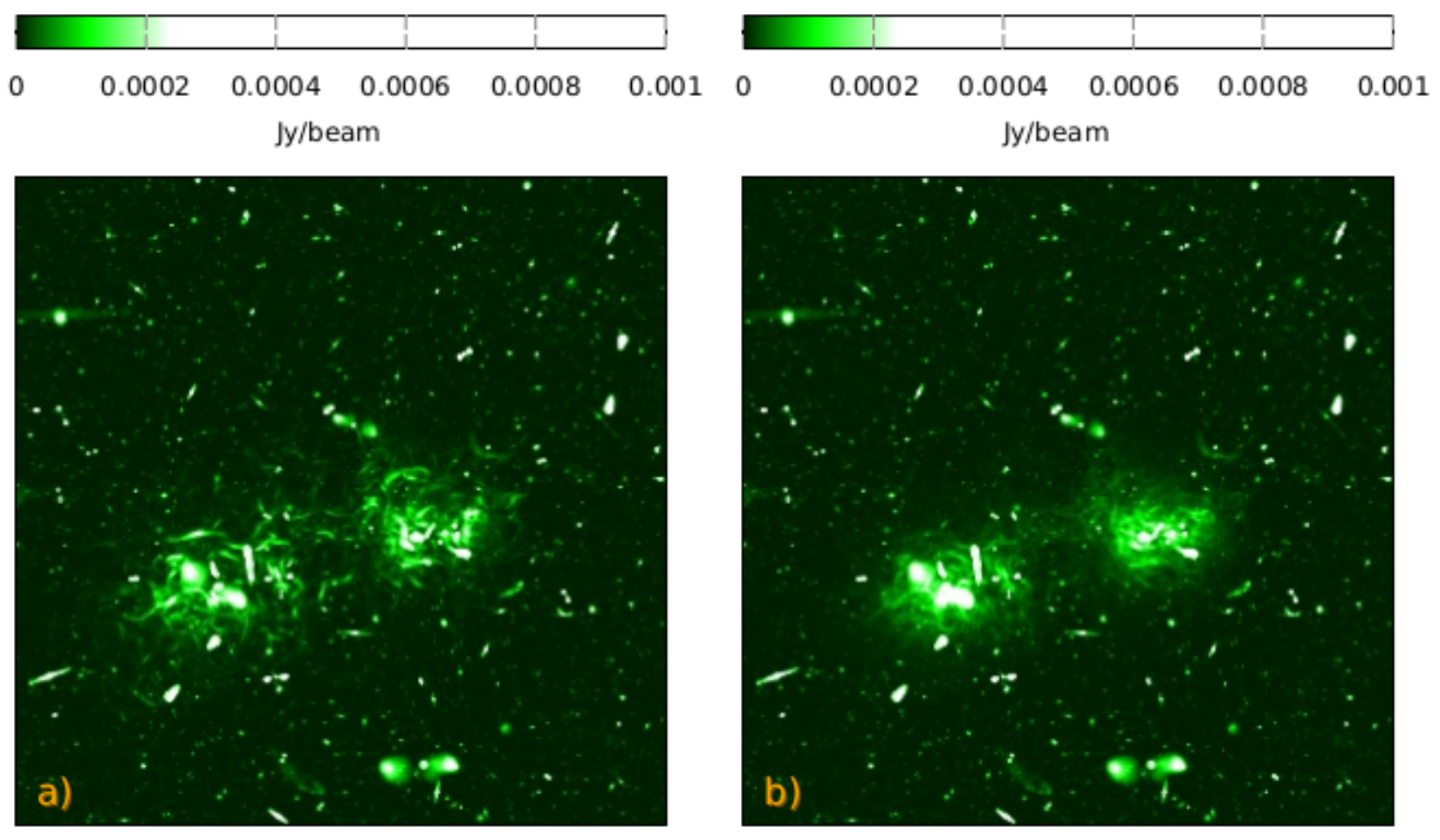}
    \caption{1.4\,GHz total intensity synthetic SKA1-MID images obtained by degrading the resolution of the simulated galaxy clusters (see Fig. \ref{fig:i_full}) to 10$\arcsec$ and adding a 0.1\,$\muup$Jy/beam noise. The cluster radio haloes are (a) produced assuming the equipartition condition, and (b) assuming a relativistic particle energy density equal to 0.3\% of the thermal plasma energy density. The surface brightness in the images goes from 0 up to $\sim$0.16mJy/beam.}
    \label{fig:img_noise}
\end{figure*}
\subsection{Synthetic radio images}
Fig. \ref{fig:img_noise} shows the synthetic radio images of the two simulated clusters produced from a data cube spanning the frequency band 950-1760\,MHz with a channel resolution of 1\,MHz. These synthetic cubes correspond to the simulated cubes of the previous subsection convolved with a Gaussian function having FWHM=10$\arcsec$, with the addition of a thermal noise of 0.1$\muup$Jy/beam.
This corresponds to the sensitivity level of $Q$ and $U$ Stokes images achievable with the SKA1-MID with an integration time of 60 hours. 
It is worth to specify that in total intensity the confusion noise is significantly higher \citep[$\sigma_c=4\,\muup$Jy/beam according to][]{loi19} and it is reached after few minutes of observing time. In practice, the actual noise in these images is therefore dominated by the confusion noise. Fig. \ref{fig:img_noise} shows the 1.4\,GHz total intensity synthetic SKA1-MID images with a) radio haloes produced assuming the equipartition condition, and b) a relativistic particles energy density equal to 0.3\% of the thermal plasma energy density. Even in these images the pixel range has been reduced but the surface brightness can be as high as 0.16\,mJy/beam. \\
The radio haloes shows a mean surface brightness computed in a circle of 1\,Mpc in radius of $\sim$40\,$\muup$Jy/beam. The filamentary structure of radio haloes in equipartition is still visible. On the other hand, the radio haloes simulated coupling thermal and non-thermal particle present a smoother morphology. 
These results highlight how high resolution and high sensitivity observations performed with next generation instruments such as the SKA1-MID could help us in understanding the nature of relativistic particles in cluster radio haloes, which is still an open and debated question in the literature.

\section{RM-synthesis application}
The $RM$-synthesis is one of the most used techniques to recover polarized signals and their associated Faraday depths in spectro-polarimetric observations. Following the formalism adopted in \citet{brent}, we will start by going over some basic concepts. We will follow by explaining the method chosen to interpret the output data.
\subsection{Basics and set-up}
Each line-of-sight where one or more polarized signals cross a magneto-ionic medium is characterised by the so-called Faraday dispersion function $F(\phi)$, which represents the emitted polarization as a function of the Faraday depth $\phi$. With $RM$-synthesis, it is possible to retrieve an approximate reconstruction of this function $\tilde{F}(\phi)$ which corresponds to the convolution of $F(\phi)$ and the $RM$ Transfer Function $R(\phi)$ (RMTF) defined as:
\begin{equation}
    R(\phi)=\frac{\int_{-\infty}^{+\infty} W(\lambda^2) e^{-2i\phi\lambda^2} d\lambda^2}{\int_{-\infty}^{+\infty} W(\lambda^2) d\lambda^2},
\end{equation}
where the sampling function or weight function $W(\lambda^2)$ is non-zero at the measured $\lambda^2$ and zero elsewhere. Indeed, the approximate reconstruction of $F(\phi)$ is obtained by Fourier transforming the observed polarized intensity as a function of $\lambda^2$, $\tilde{P}(\lambda^2)$:
\begin{equation}
    \tilde{F}(\phi)=\frac{\int_{-\infty}^{+\infty} \tilde{P}(\lambda^2) e^{-2i\phi\lambda^2} d\lambda^2}{\int_{-\infty}^{+\infty} W(\lambda^2) d\lambda^2}.
    \label{eq:rmsynth}
\end{equation}
All these functions are complex, and $\tilde{P}(\lambda^2)=Q(\lambda^2)+iU(\lambda^2)$.
Eq. \ref{eq:rmsynth} corresponds to a de-rotation of the $Q$ and $U$ Stokes parameters, assuming a given value of Faraday depth in a given interval, and performed channel by channel, pixel by pixel. The resulting $Q$, and $U$ Stokes parameters cubes as a function of the Faraday depth can be used to build a polarized intensity cube as shown in Eq. \ref{eq:p} by taking the modulus of the Faraday dispersion function. \\
The accuracy of the $RM$-synthesis technique is determined by the $\lambda^2$ coverage $\Delta\lambda^2$, the minimum wavelength squared $\lambda_{\rm min}^2$, and the channel width $\delta\lambda^2$. In particular, their values define three important parameters: a) the resolution in Faraday space (which corresponds to the FWHM of the RMTF), b) the largest scale in Faraday depth, and c) the maximum observable Faraday depth to which the technique is sensitive. For a bandwidth between 950 and 1760\,MHz with a frequency channel of 1\,MHz, the former parameters become:
\begin{eqnarray}
    \delta \phi &\approx& \frac{2\sqrt{3}}{\Delta \lambda^2} = 49\,\rm rad/m^2\\ \nonumber
    \rm max-scale &\approx& \frac{\pi}{\lambda_{\rm min}^2} = 108\,\rm rad/m^2 \\ \nonumber
    |\phi_{\rm max}| &\approx& \frac{\sqrt{3}}{\delta\lambda^2} \approx 24000\,\rm rad/m^2 ,
\end{eqnarray}
where $\delta\lambda^2$ is computed at the centre of the bandwidth. 
This observational set-up is a compromise between minimising the RMTF FWHM (and therefore Faraday space resolution), which improves at lower frequencies, and the need for sensitivity to large scale structures in Faraday space, which improves at high frequencies. Indeed, with a frequency band between 350 and 1050\,MHz the resolution and the maximum scale are respectively 5\,rad/m$^2$ and 38\,rad/m$^2$. For example, a radio halo filament having a width of 50\,kpc located in the plane of the sky with a thermal density of 0.006\,cm$^{-3}$ and a magnetic field of 0.5\,$\muup$G would produce a Faraday thick region of $\sim$120\,rad/m$^2$. This quantity roughly corresponds to the maximum scale measurable in Faraday depth between 750 and 1760\,MHz while it is three times the maximum scale between 350 and 1050\,MHz.\\
The Faraday space in this work is sampled with a cell size of 10 rad/m$^2$ between -2000 and 2000 rad/m$^2$, in order to have $\sim$5 pixels in the RMTF FWHM. The weight function it is assumed to be equal to 1 for each observed $\lambda^2$ and zero elsewhere. 

\subsection{Interpretation of the products}
As mentioned in the previous Sections, the total intensity images of this work, unlike the polarized intensity ones, are confusion limited. Therefore, there will be polarized sources above the noise level in polarization which are not distinguishable from the noise in the total intensity images. The usual approach is to perform a cut on the $RM$-synthesis output images by creating a 3$\sigma_I$ mask with the total intensity map but this will significantly limit the information potentially available in polarization. 
Another issue related to the interpretation the $RM$-synthesis results is the question of obtaining the polarized intensity from the cubes. Since there could be, especially in the case of radio haloes, many polarized emissions along the line-of-sight, the result is that the Faraday dispersion function is likely to feature more than one peak. The resulting polarization is then the sum of the contribution along the spectrum. However, the polarized intensity cube contain a positive {\it bias} due to the fact that it is the square root of the sum of two squares ($||P||=\sqrt{Q^2+U^2}$) which produces a mean $||P||>0$ even in absence of source signals because of the presence of the thermal noise. The plateau in the polarization image obtained by summing up all the polarization as a function of the Faraday depth can be significantly high as it propagates the positive bias a number of times equal to the Faraday depth channels, resulting in a very low S/N ratio.\\
In order to avoid the emergence of the positive bias and the loss of polarized sources above the 3$\sigma_I$-level in polarization, in this work the polarized intensity cube is treated as follows:
\begin{itemize}
    \item first, a mean is evaluated from the first channel of the polarized intensity cube as a function of the Faraday depth where $\phi=-2000\,\rm rad/m^2$ and it is unrealistic to have a real cluster $RM$. Indeed, to obtain such values, for a thermal density of $\sim10^{-3}\,\rm cm^{-3}$ the magnetic field should be of the order of few mG, which is three orders of magnitudes larger than the typical intracluster magnetic field strength. Therefore, it is possible to evaluate a mean over the full image in polarization at $\phi=-2000\,\rm rad/m^2$ and subtract it from each pixel of the polarized intensity cube as a function of the Faraday depth. The results is that the spectra along $\phi$ will scatter around zero and the sum of the polarization contributions along $\phi$ will not be affected by the positive bias.
    \item Then, the polarized image is created by summing up all the contribution along $\phi$ of the polarized intensity cube corrected for the bias.
    \item The value corresponding to the maximum peak in polarized intensity is considered to be the measured $RM$. This is clearly an arbitrary choice, and there is no guarantee that it is the correct one for a given line-of-sight. Indeed, the cluster $RM$ could be the Faraday depth associated to a lower polarized emission. A choice of some kind, however must be made: this is one of the most important issues in the interpretation of this technique and it will be the subject of a future work. The resulting $RM$ images are multiplied by a factor $(1+z)^2$, with $z$ the cluster redshift to correct for the Doppler effect on the observed frequencies.
    \item As a final step, it is necessary to filter the polarization and the $RM$ images, selecting the pixels where the information is related to a real signal. To do this in the simulated images, we create a mask where the polarized intensity is different from zero and we used this mask to blank the $RM$ images. In the synthetic images, the procedure is not so  straightforward, because of the presence of noise. We thus use an iterative process to evaluate the Root Mean Square (rms) value in a region of the polarization image where no obvious sources are present, clipping the pixels above 3 times the rms. The mask is then created by selecting all pixel with signal 4 times above rms. The factor 4 has been chosen as a compromise between the need to discard noise signals and the need to preserve as much pixels as possible. 
\end{itemize}

 \begin{figure*}
    \centering
    \includegraphics[width=0.78\textwidth]{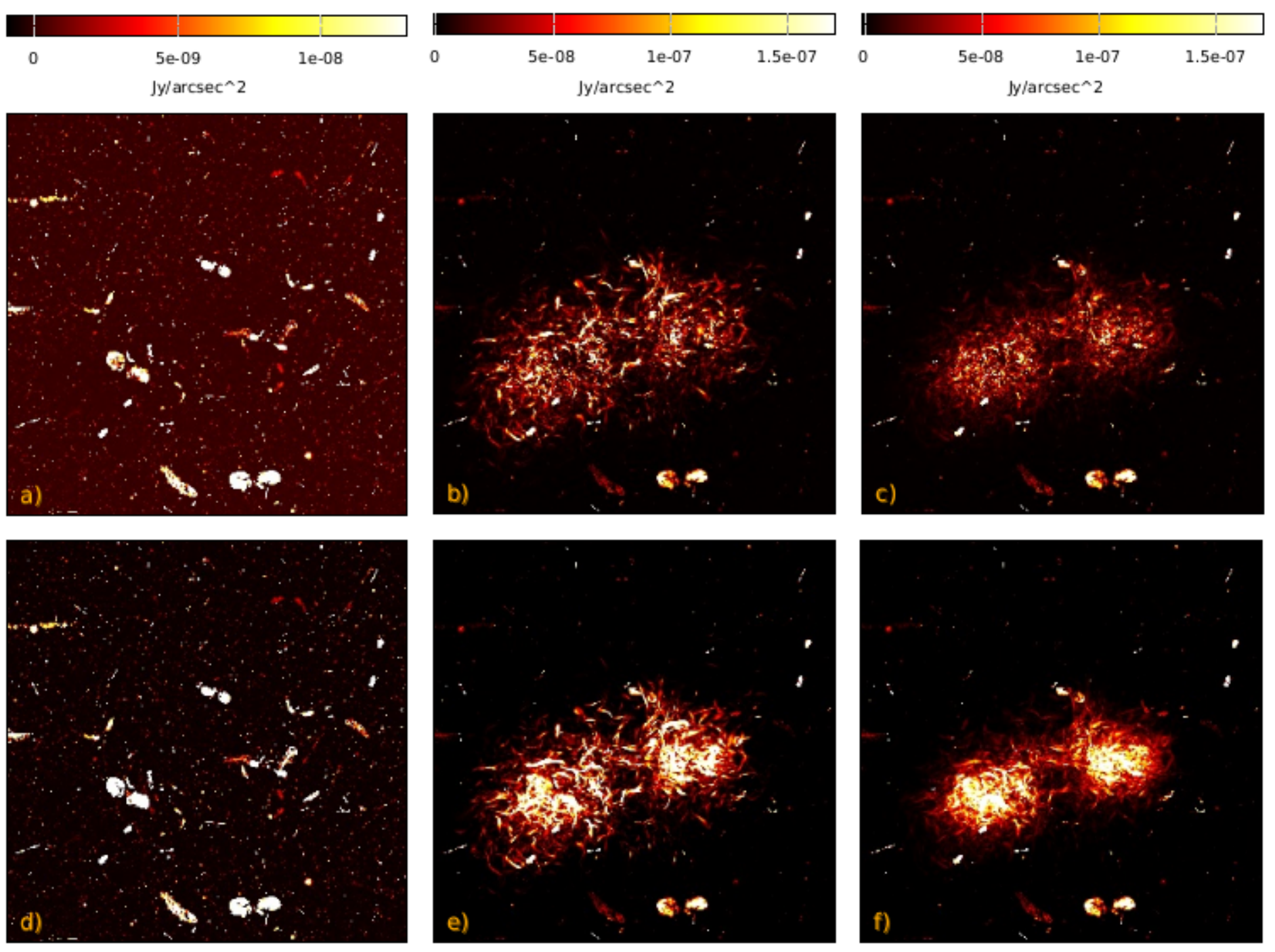}
    \caption{1.4 GHz linearly polarized simulated images. First row shows the linearly polarized intensity computed assuming $RM$=0, second row the linearly polarized intensity reproduced after the application of the $RM$-synthesis. From left to right: clusters without radio haloes, with radio haloes under the equipartition condition, and simulated coupling between thermal and non-thermal particles.}
    \label{fig:comp_p}
\end{figure*}
\begin{figure*}
    \centering
    \includegraphics[width=0.78\textwidth]{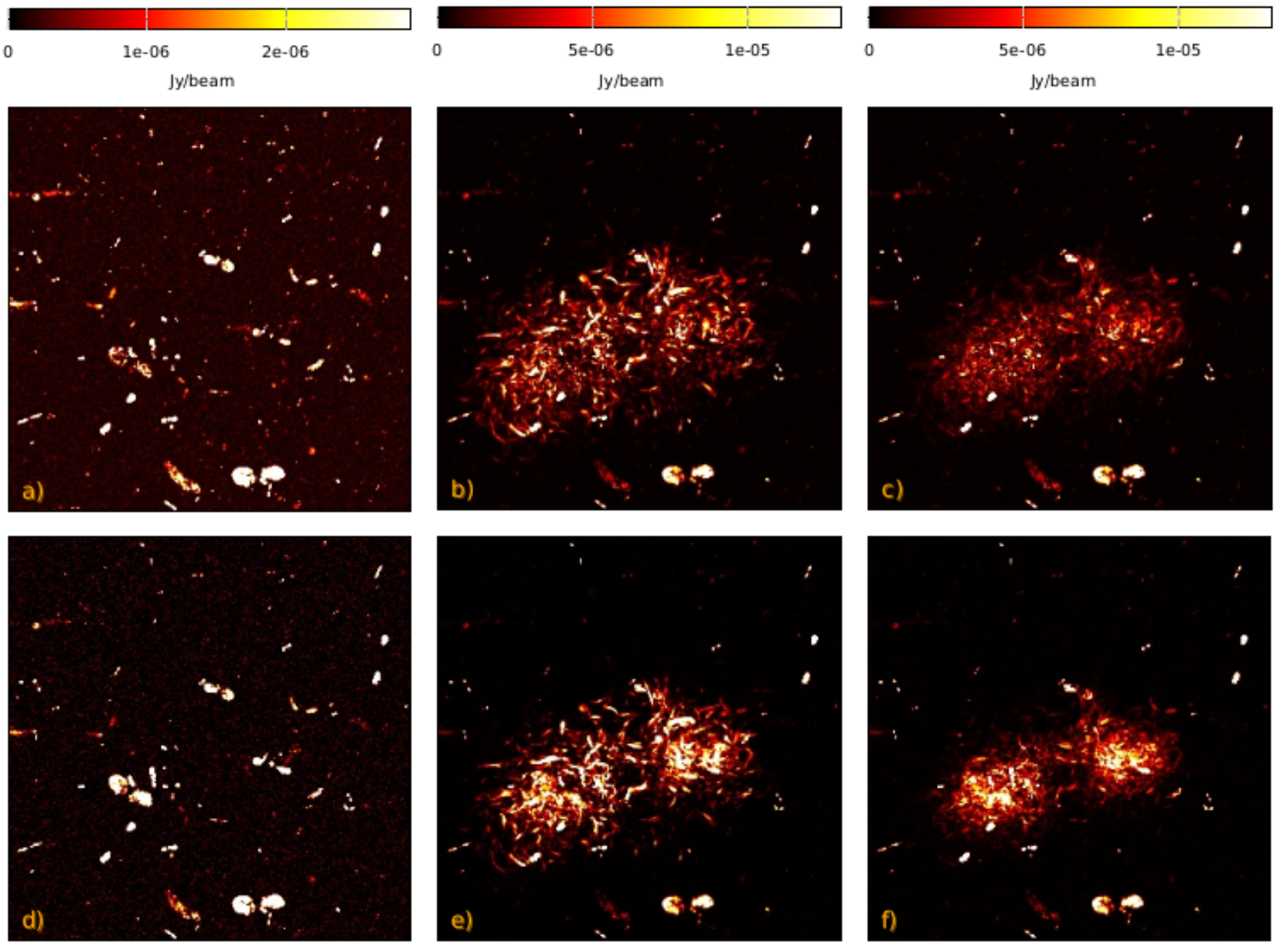}
    \caption{1.4 GHz linearly polarized synthetic images. See caption of Fig. \ref{fig:comp_p} for more details.}
    \label{fig:comp_p_noise}
\end{figure*}
\section{RM-synthesis results}
\label{sect:rmres}
In this Section, we show the results of the procedure described  previously, focusing on linearly polarized emission and the $RM$ images obtained from simulated and synthetic data.

\subsection{Polarization results}
\label{sect:p}
Figs. \ref{fig:comp_p} and \ref{fig:comp_p_noise} show the result of the $RM$-synthesis technique on simulated and synthetic data. 
First rows show the observed linearly polarized intensity computed assuming a $RM$ equal to zero, while second rows show the linearly polarized intensity retrieved after the application of the $RM$-synthesis. From left to right the panels refer to clusters without radio haloes, with radio haloes under the equipartition condition, and assuming coupling between non-thermal and thermal particle energy density. \\
Simulated images of clusters without radio haloes show the presence of a large number of background sources which are below the noise level in the synthetic ones. This is mainly due to beam depolarization, which reduces the polarized signal of the faint background sources. Along with the presence of thermal noise, beam depolarization prevents the recover of background signals in the synthetic images. Indeed, the radio sources near to the cluster centres in these images are cluster radio sources. We discuss the consequence of a very low background contribution in Section \ref{sect:rman}.\\
As for the total intensity emission, the linearly polarized emission suggests a quite different morphology of the two kind of simulated radio haloes: radio haloes in equipartition (b and e) shine in polarization even at large distances from the cluster centres, showing a discontinuous arrangement of bright and faint filaments, while radio haloes with coupling between thermal and non-thermal particles (c and f) show a clear and regular decrease of the surface brightness from the centre to the outskirts and the filamentary morphology is less pronounced.
As expected, the polarized intensity from $RM$-synthesis is larger than what computed assuming $RM$=0, both in the case of simulated and synthetic clusters, suggesting that bandwidth depolarization is a significant effect here. This is more evident in the central region of the clusters, where the higher values of $RM$ are responsible of the strong depolarization of $\sim$20-30\% toward the centre for simulated images, and of $\sim$40-60\% for synthetic ones. However, comparing between simulated and synthetic results, we can see that some polarized emission in synthetic data, especially that of the radio haloes in the outskirts, is lost due to the fact that this emission does not emerge from the noise in the $RM$-synthesis cubes.
\begin{figure*}
    \centering
    \includegraphics[width=0.78\textwidth]{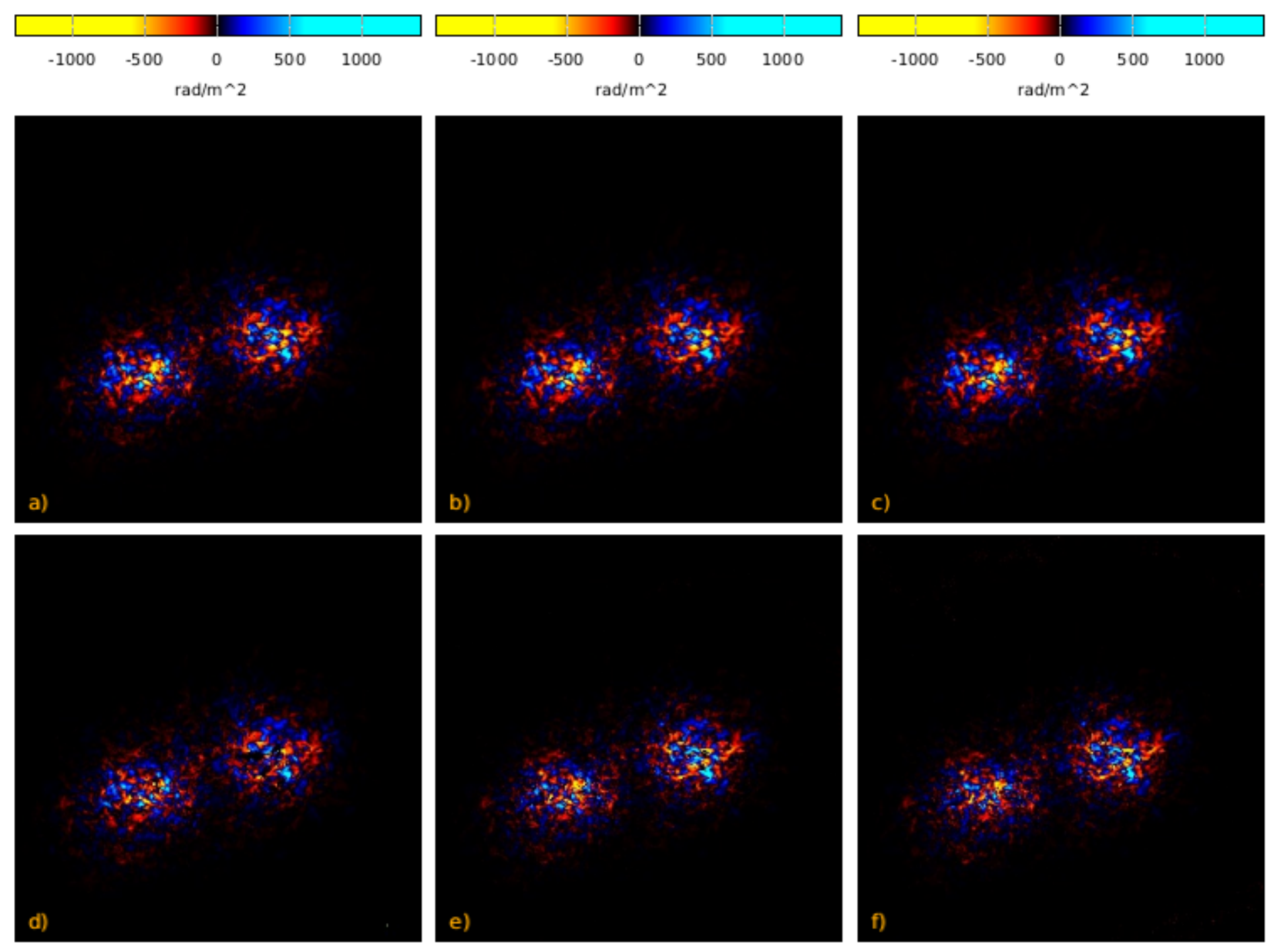}
    \caption{RM simulated images. First and second rows refer respectively to the intrinsic $RM$ (computed from the MHD cubes) and to the $RM$ retrieved after the application of the $RM$-synthesis. From left to right, images refers to the clusters without radio haloes, radio haloes in equipartition, and simulated coupling between thermal and non-thermal particles.}
    \label{fig:rmsynth_img}
\end{figure*}
Finally, it is important to notice how the presence of a large scale diffuse source, i.e. the radio halo, is crucial in sampling the $RM$ over a large area of the cluster: this will be possible with data acquired with SKA1-MID and analysed with the $RM$-synthesis technique.

\begin{figure*}
    \centering
    \includegraphics[width=0.78\textwidth]{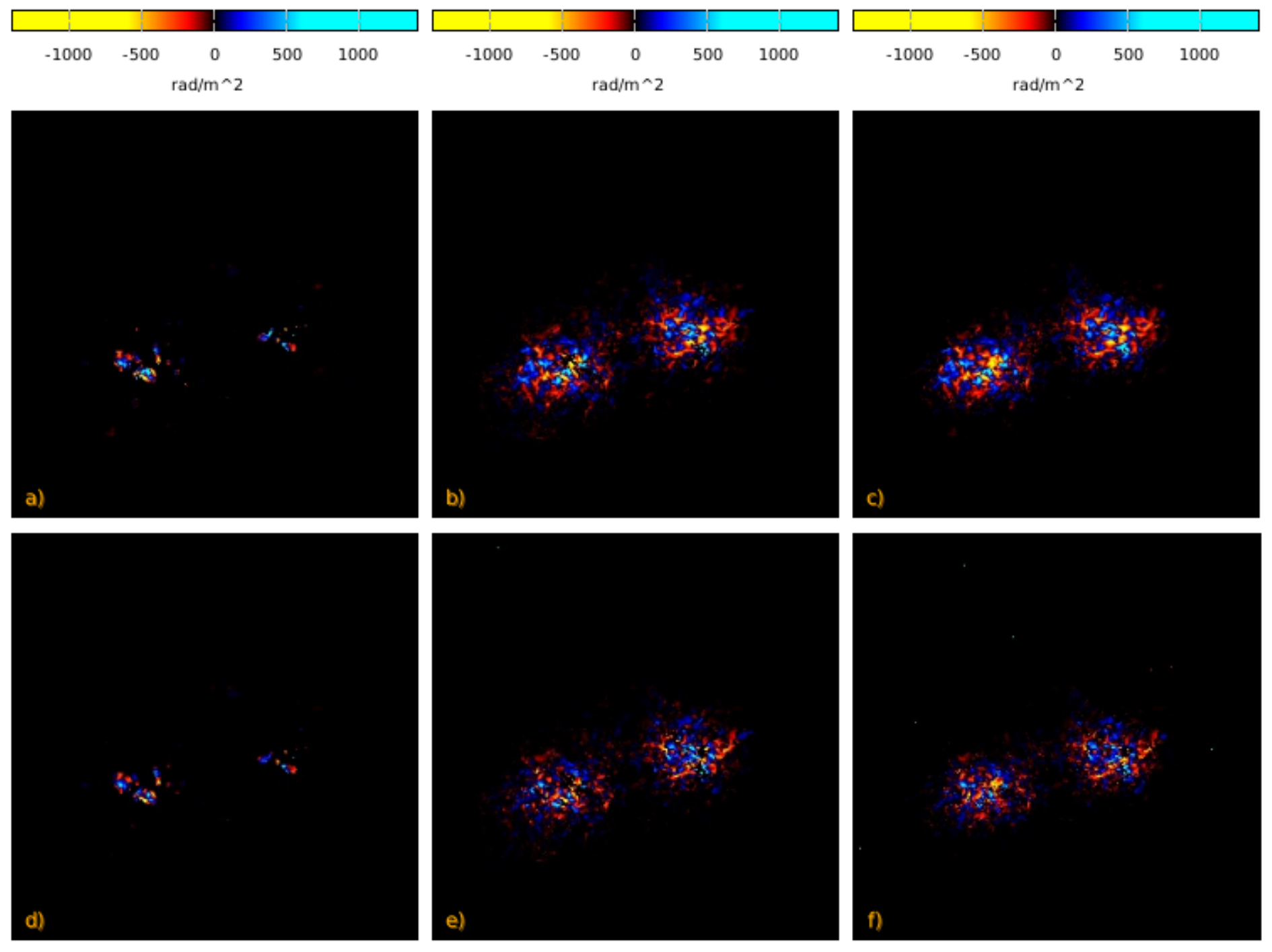}
    \caption{RM synthetic images. See caption of Fig. \ref{fig:rmsynth_img} for more details.}
    \label{fig:rmsynth_img_noise}
\end{figure*}

\subsection{RM images}
Fig. \ref{fig:rmsynth_img} shows the results concerning the $RM$ for simulated and synthetic data respectively. First rows refer to the intrinsic $RM$ (computed from the MHD cubes) while second rows to the $RM$ retrieved after the application of the $RM$-synthesis. From left to right, images show the $RM$ of the clusters without radio haloes, clusters with radio haloes in equipartition, and simulated coupling between thermal and non-thermal particles. Similarly, the results for synthetic data are reported in Fig. \ref{fig:rmsynth_img_noise}. \\
\begin{figure*}
    \centering
    \includegraphics[width=0.32\textwidth]{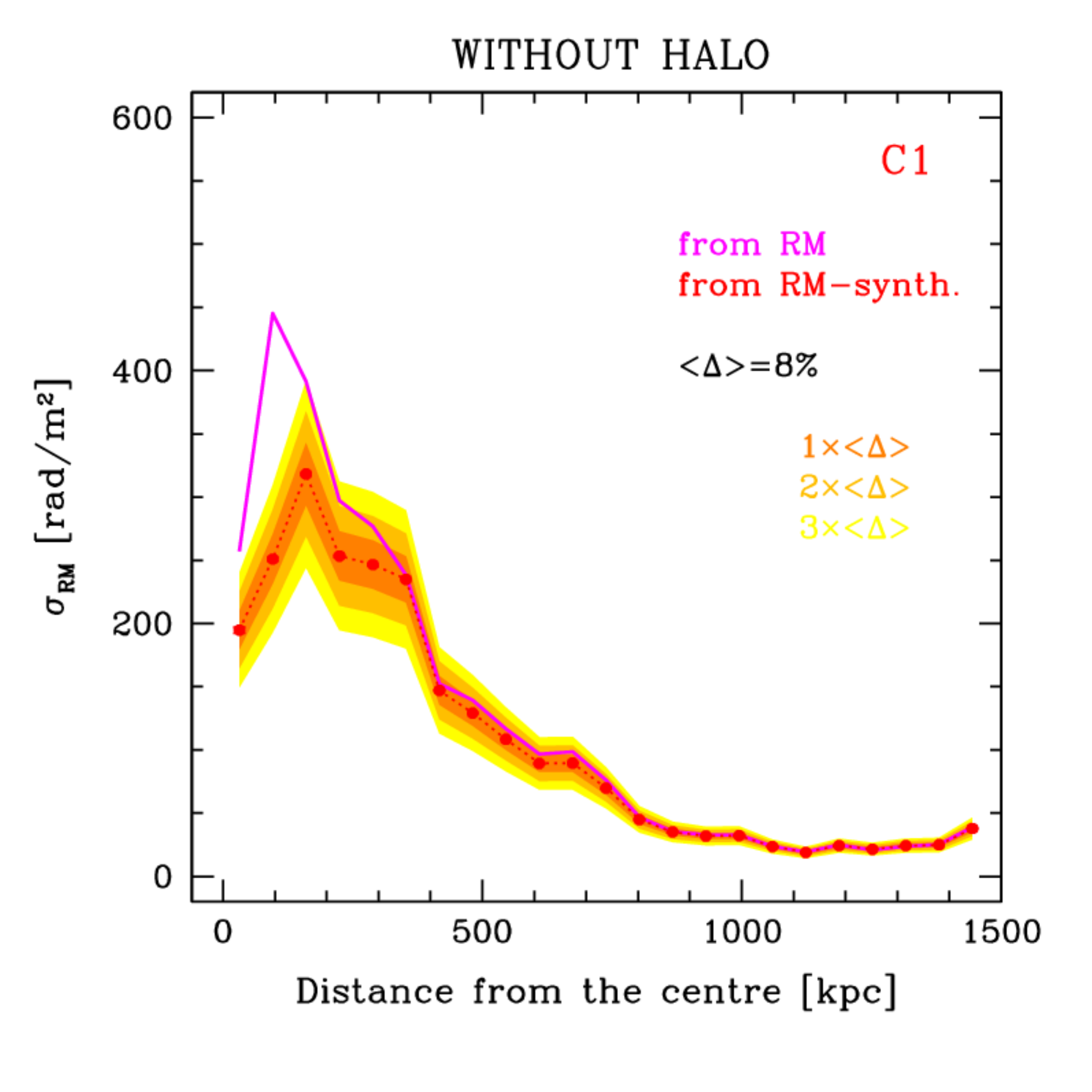}
    \includegraphics[width=0.32\textwidth]{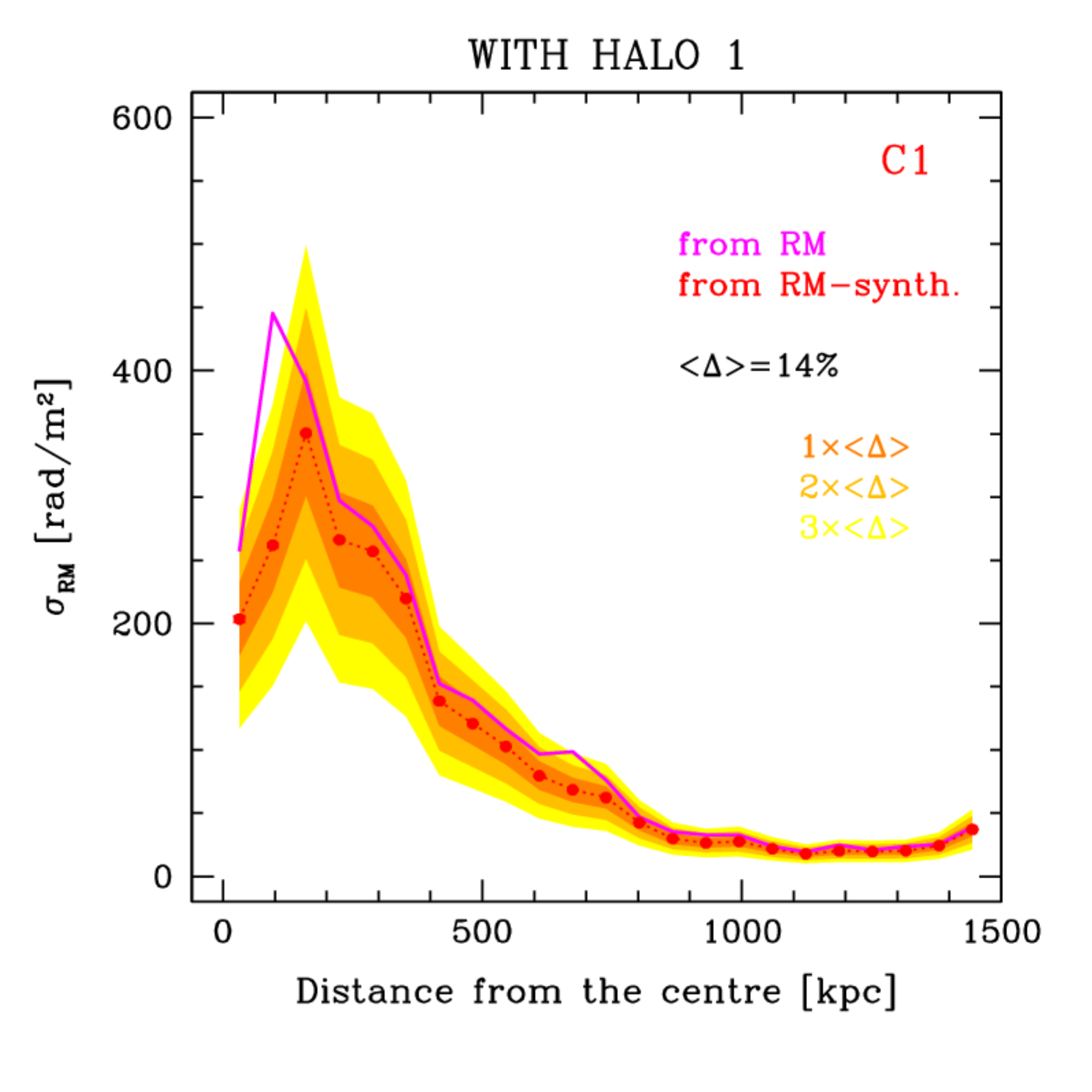}
    \includegraphics[width=0.32\textwidth]{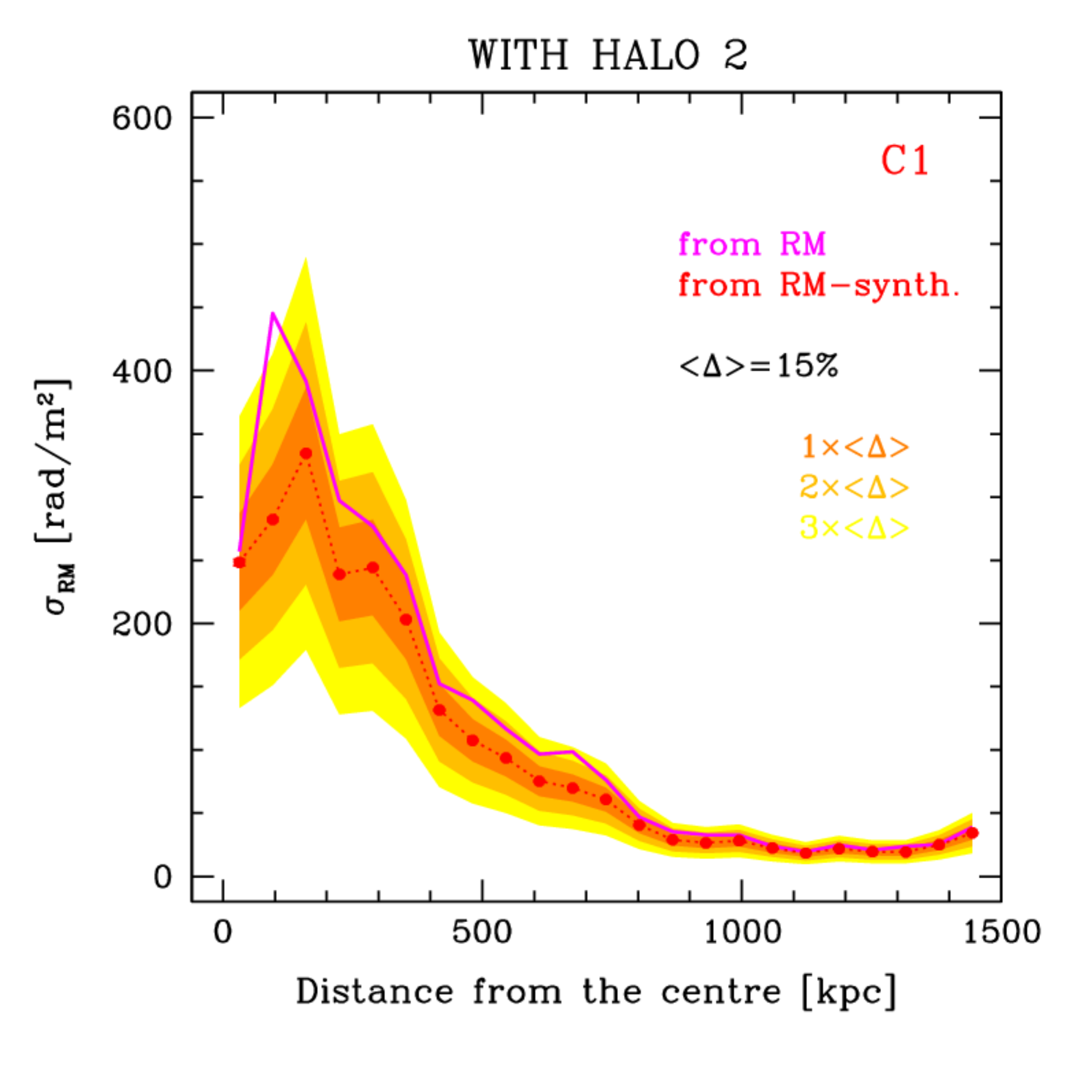}\\ 
    \includegraphics[width=0.32\textwidth]{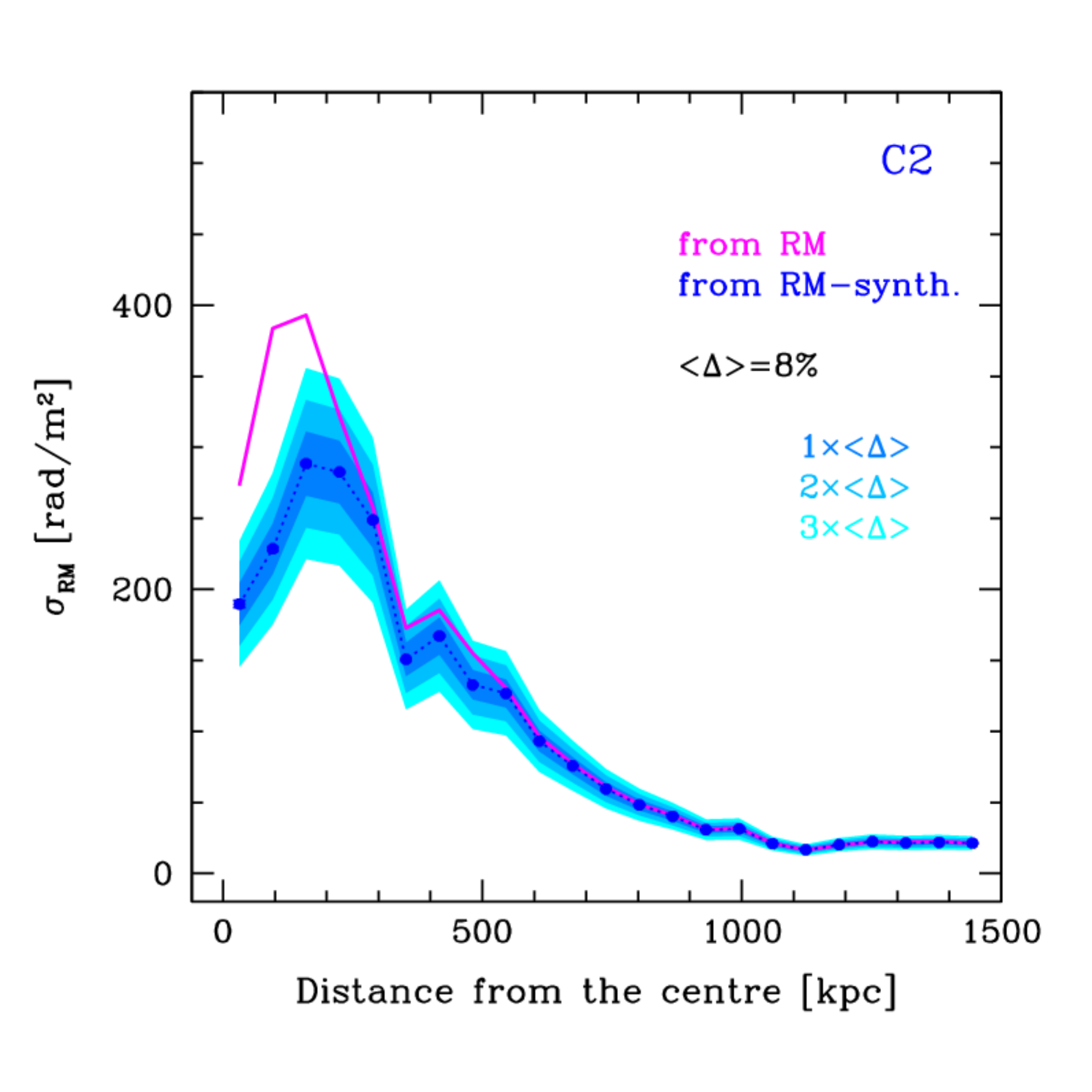}
    \includegraphics[width=0.32\textwidth]{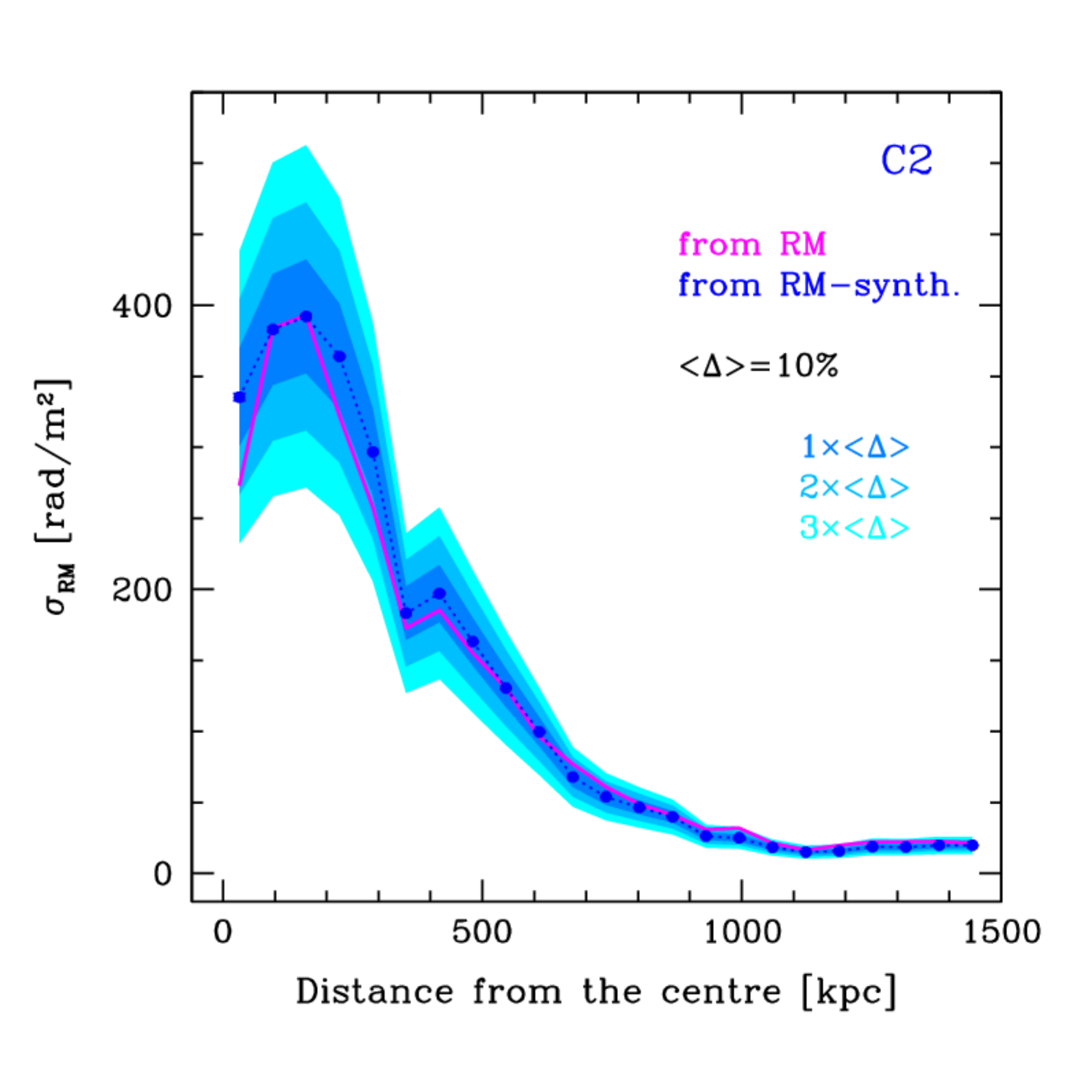}
    \includegraphics[width=0.32\textwidth]{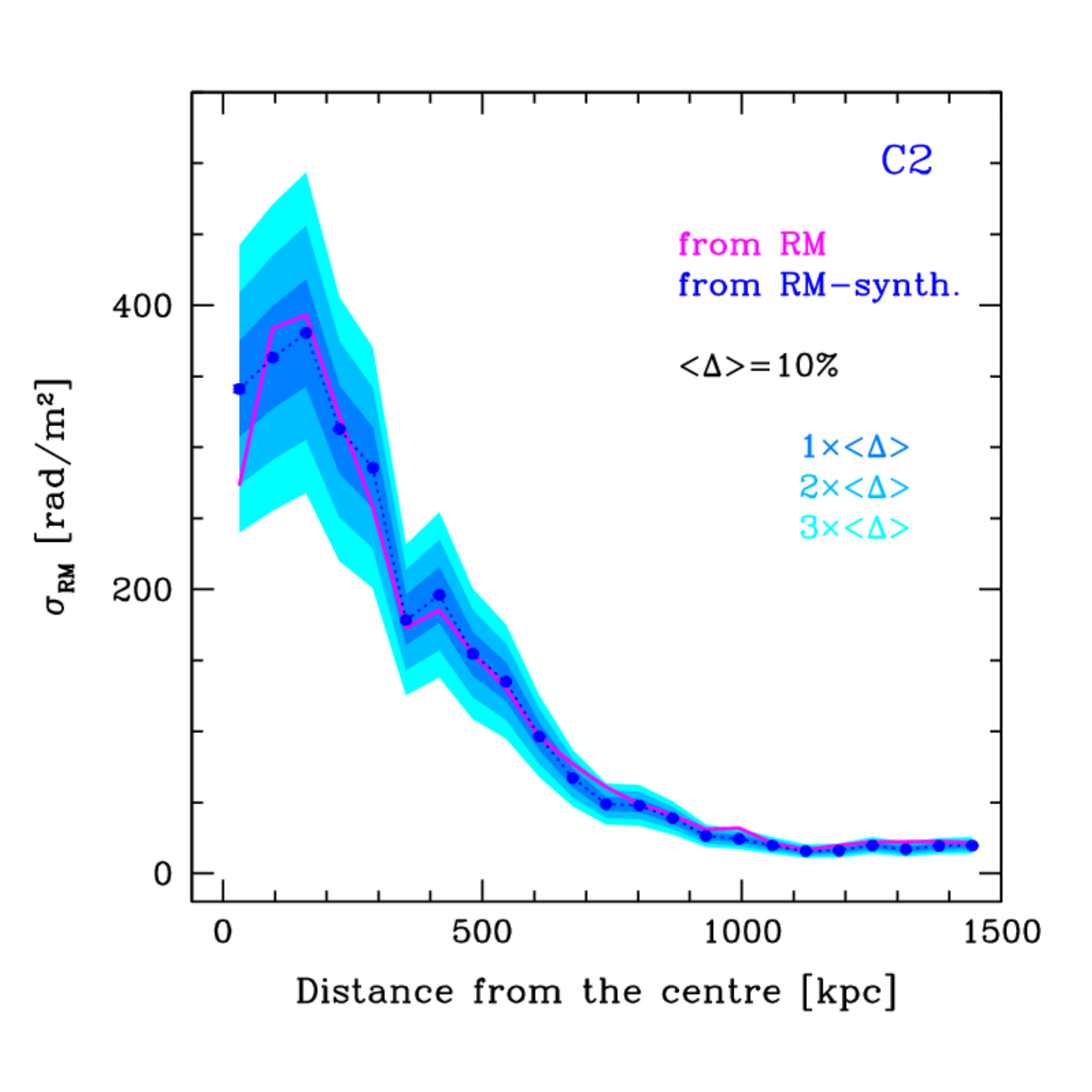}
    \caption{Simulated $\sigma_{RM}$ profiles computed in annuli of 6 pixel of width corresponding to a spatial width of $\sim$64\,kpc. Top and bottom panels refer to the C1 and C2 simulated clusters. From left to right, there are shown the results for cluster without radio haloes, radio haloes in equipartition, and simulated by coupling the non-thermal and thermal particles. In each panel a solid magenta line represent the profile computed from the intrinsic $RM$ images (Fig. \ref{fig:rm}), and points and the dotted red lines refer to the profile computed from the output Faraday depth (bottom images of \ref{fig:rmsynth_img}). $\langle \Delta \rangle$ indicate the average difference in percentage between the results of the $RM$-synthesis and of the intrinsic $RM$. Shadow region are traced at 1, 2, and 3$\times\langle\Delta\rangle$.}
    \label{fig:rm_prof}
\end{figure*}At the cluster centres, the $RM$-synthesis $RM$ values are in general lower, in absolute value, than the intrinsic ones. The central region of clusters is sampled in a better way in the case of simulated data with respect to the synthetic data, especially for clusters without radio haloes. This is caused by the presence of numerous background sources in the simulated images which unfortunately are below the noise level in the synthetic ones. No substantial differences appear between the two kind of simulated radio haloes except for a slightly larger area sampled by the synthetic radio haloes in equipartition due the larger extension of the corresponding polarized emission.
\newline
\newline
\section{RM analysis}
\label{sect:rman}
The aim of this work is to compare the results of the $RM$-synthesis applied on SKA1-MID synthetic data with the input information and to analyse discrepancies between the two. The determination of cluster magnetic fields here is based on the $RM$ properties. Indeed, the radial profile of the $\sigma_{RM}$ and the $RM$ structure function, together with the fractional polarization, are the observables used to constrain the magnetic field power spectrum \citep[e.g.][]{murgia04,laing08,govoni06,guidetti08,guidetti10,bonafede10,vacca10,vacca12,govoni17}. 
In what follows this two different diagnostic quantities are analysed.
\begin{figure*}
    \centering
    \includegraphics[width=0.32\textwidth]{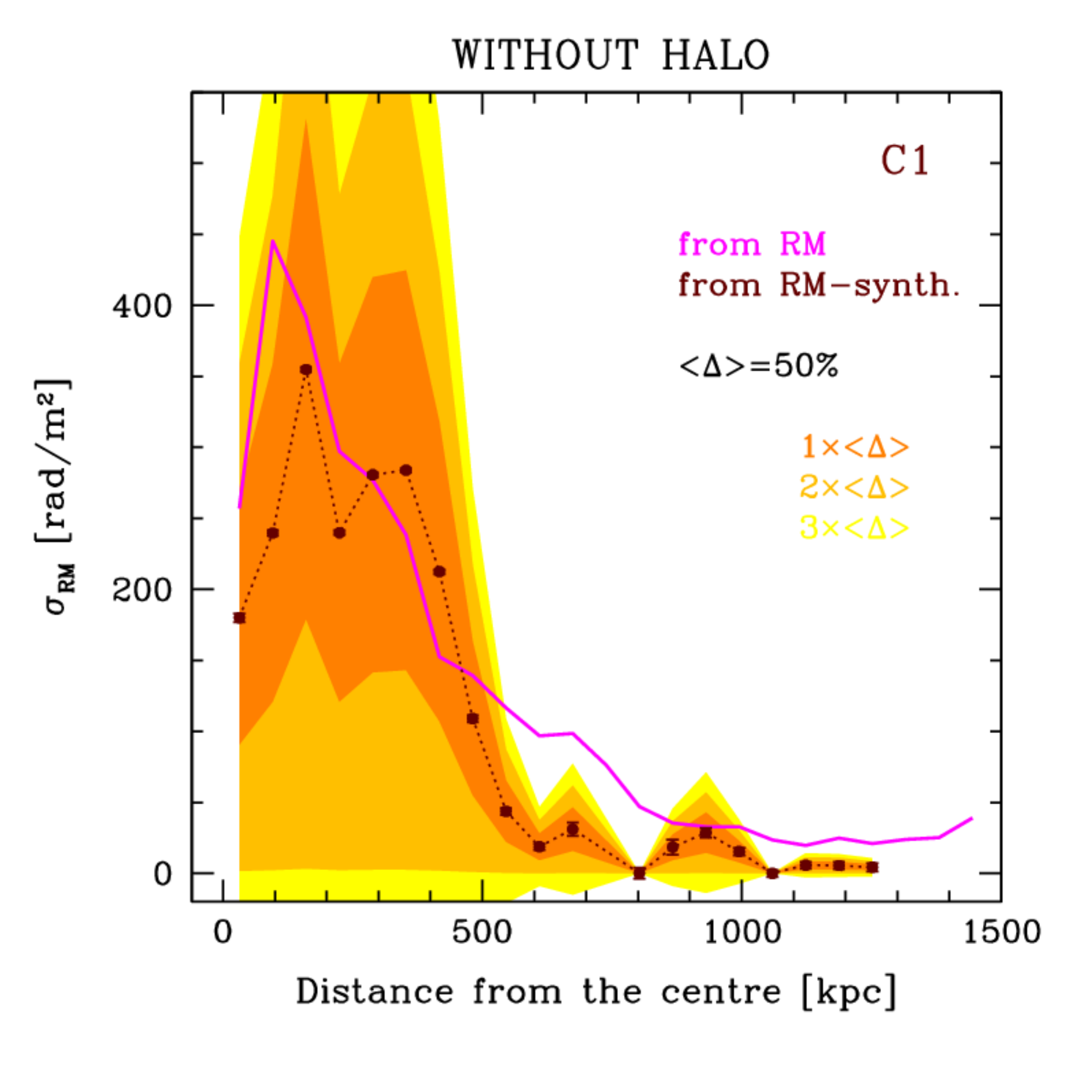}
    \includegraphics[width=0.32\textwidth]{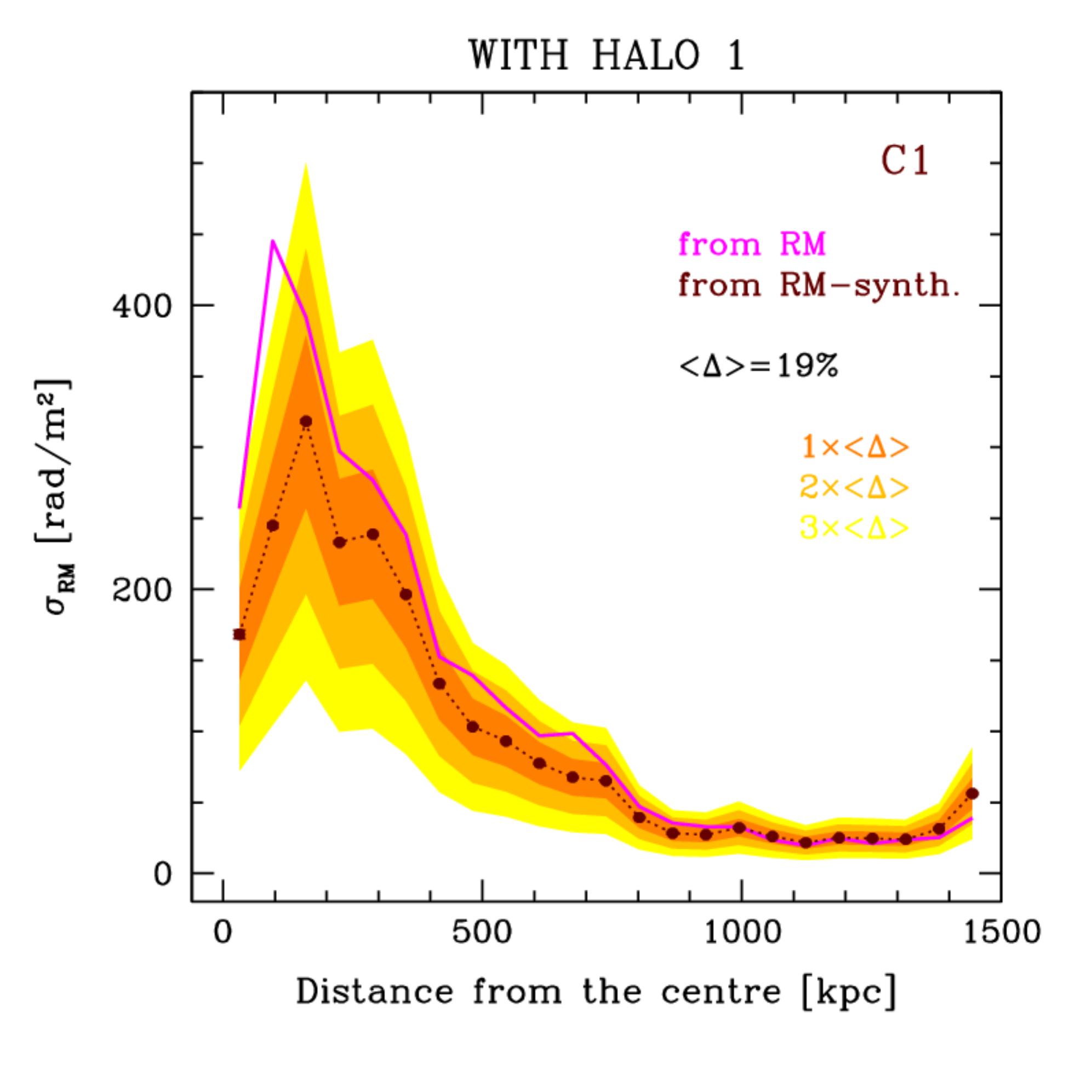}
    \includegraphics[width=0.32\textwidth]{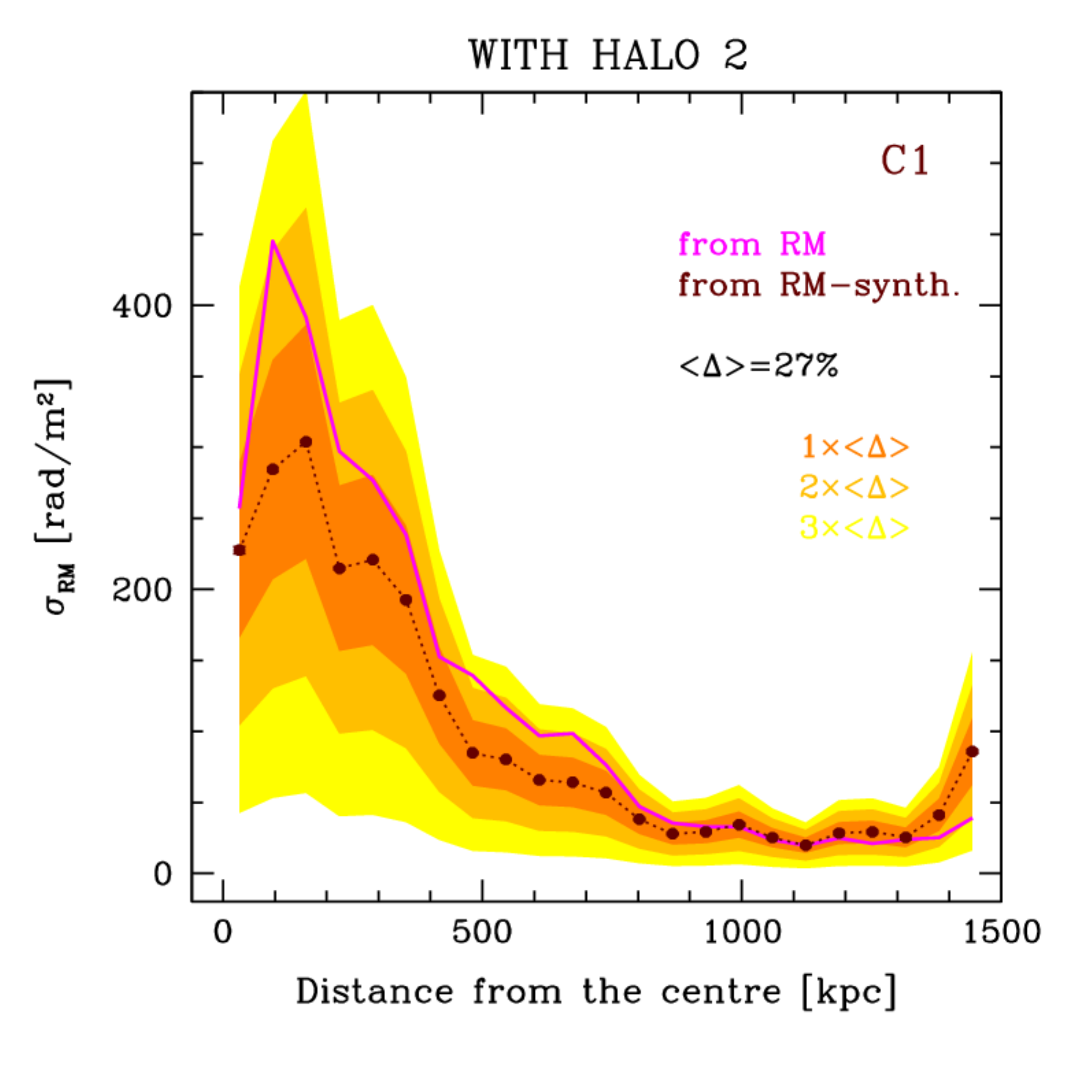}
    \includegraphics[width=0.32\textwidth]{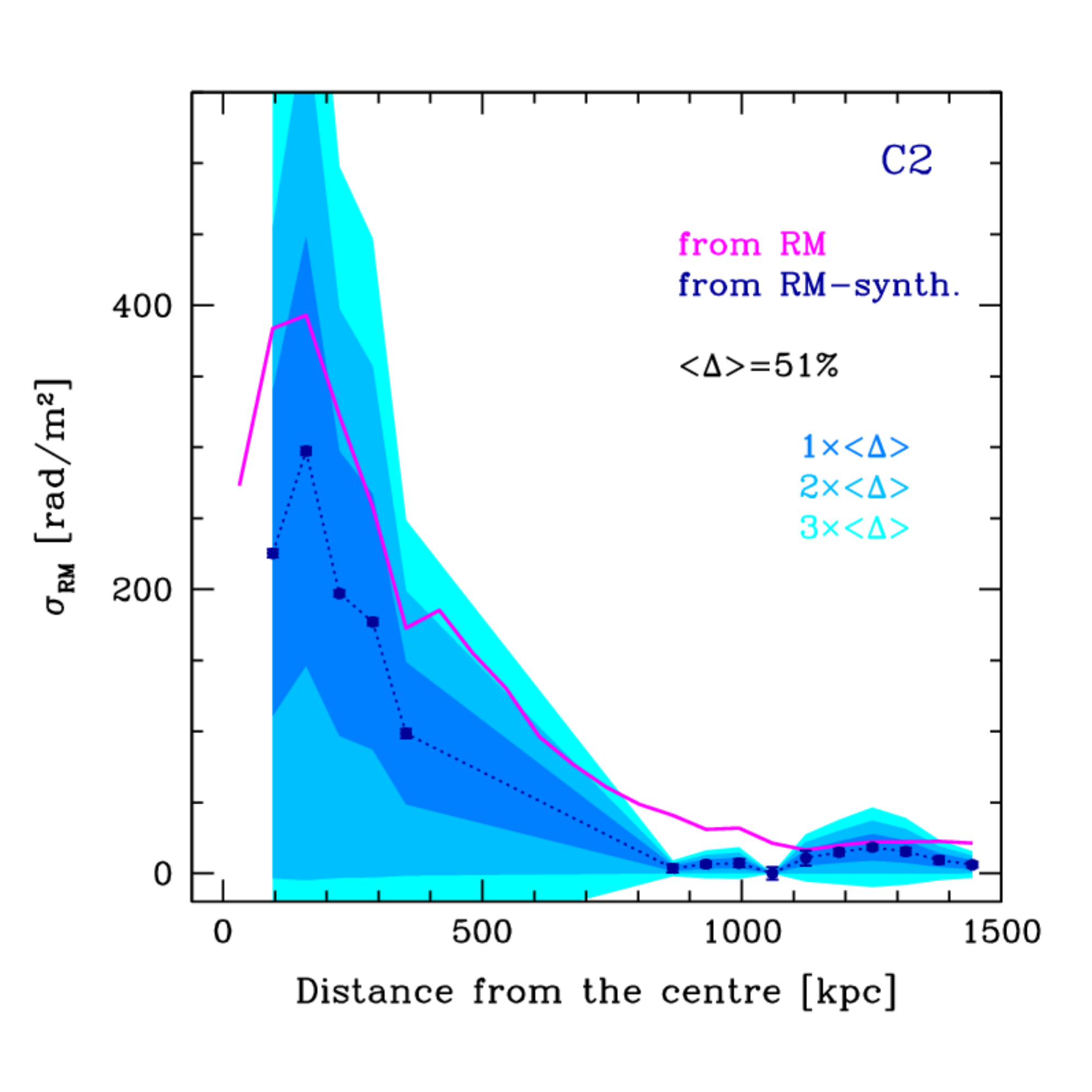}
    \includegraphics[width=0.32\textwidth]{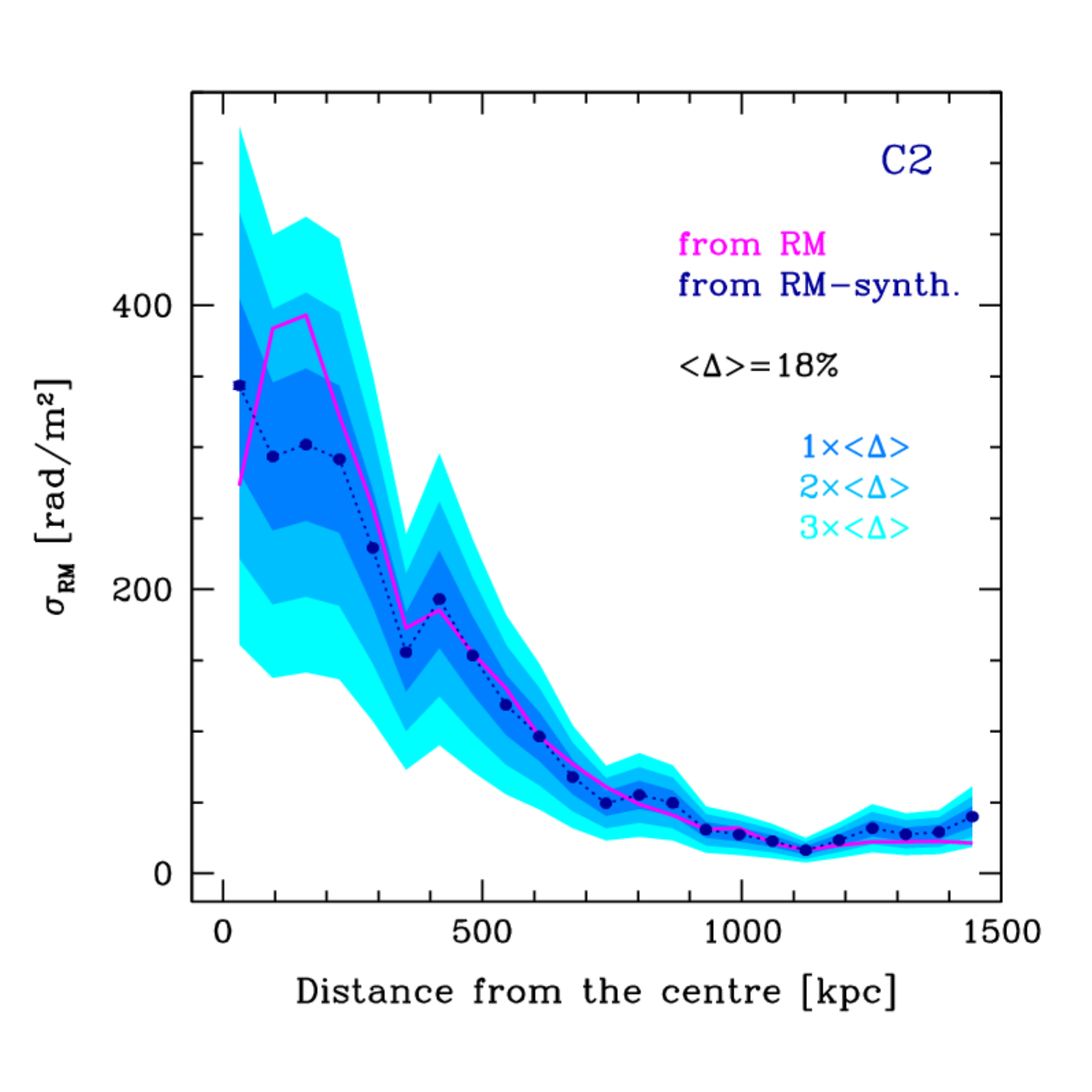}
    \includegraphics[width=0.32\textwidth]{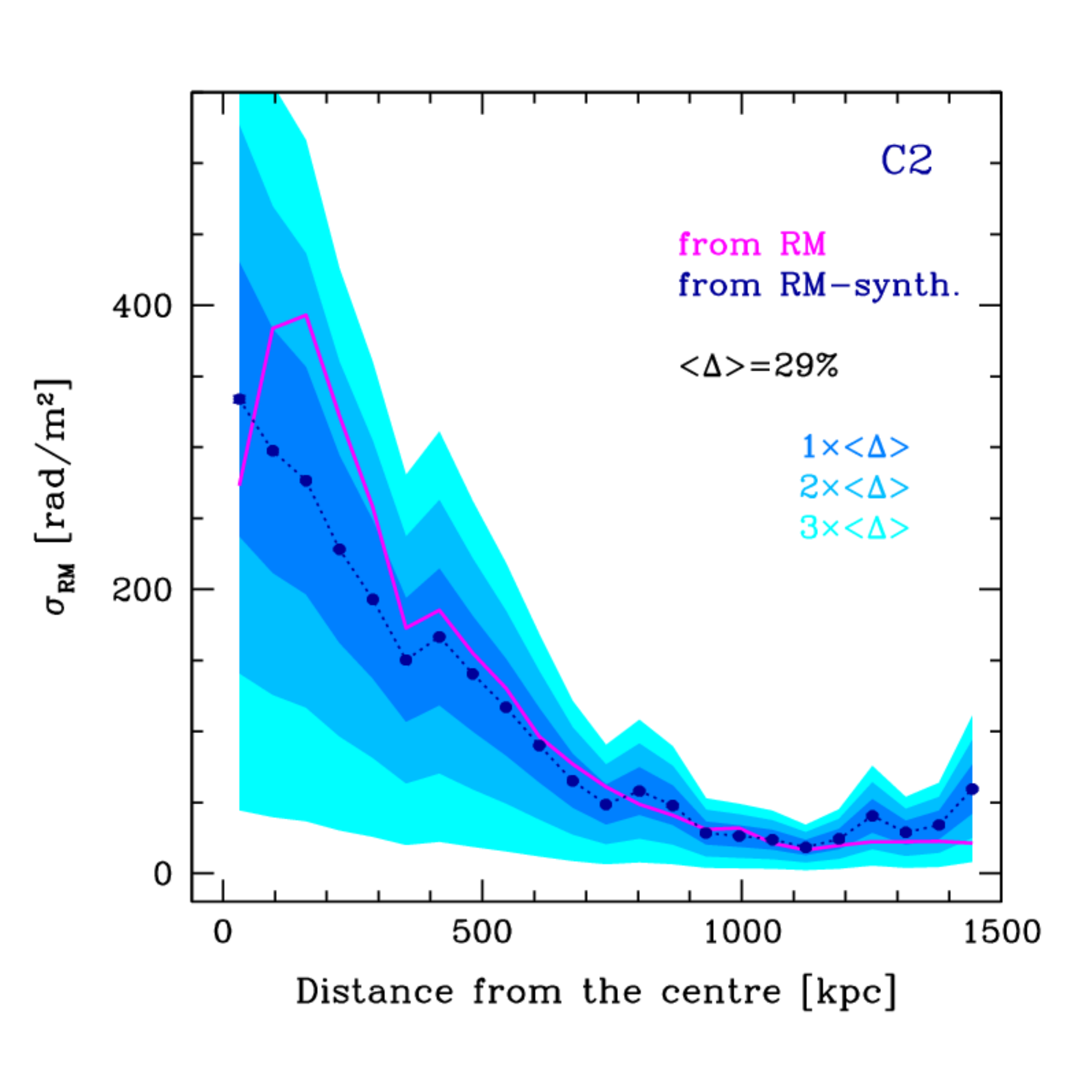}
    \caption{Synthetic $\sigma_{RM}$ profiles computed in annuli of 6 pixel of width corresponding to a spatial width of $\sim$64\,kpc. See caption of Fig. \ref{fig:rm_prof} for more details.}
    \label{fig:rm_prof_noise}
\end{figure*}
\begin{figure}
    \centering
    \includegraphics[width=0.47\textwidth]{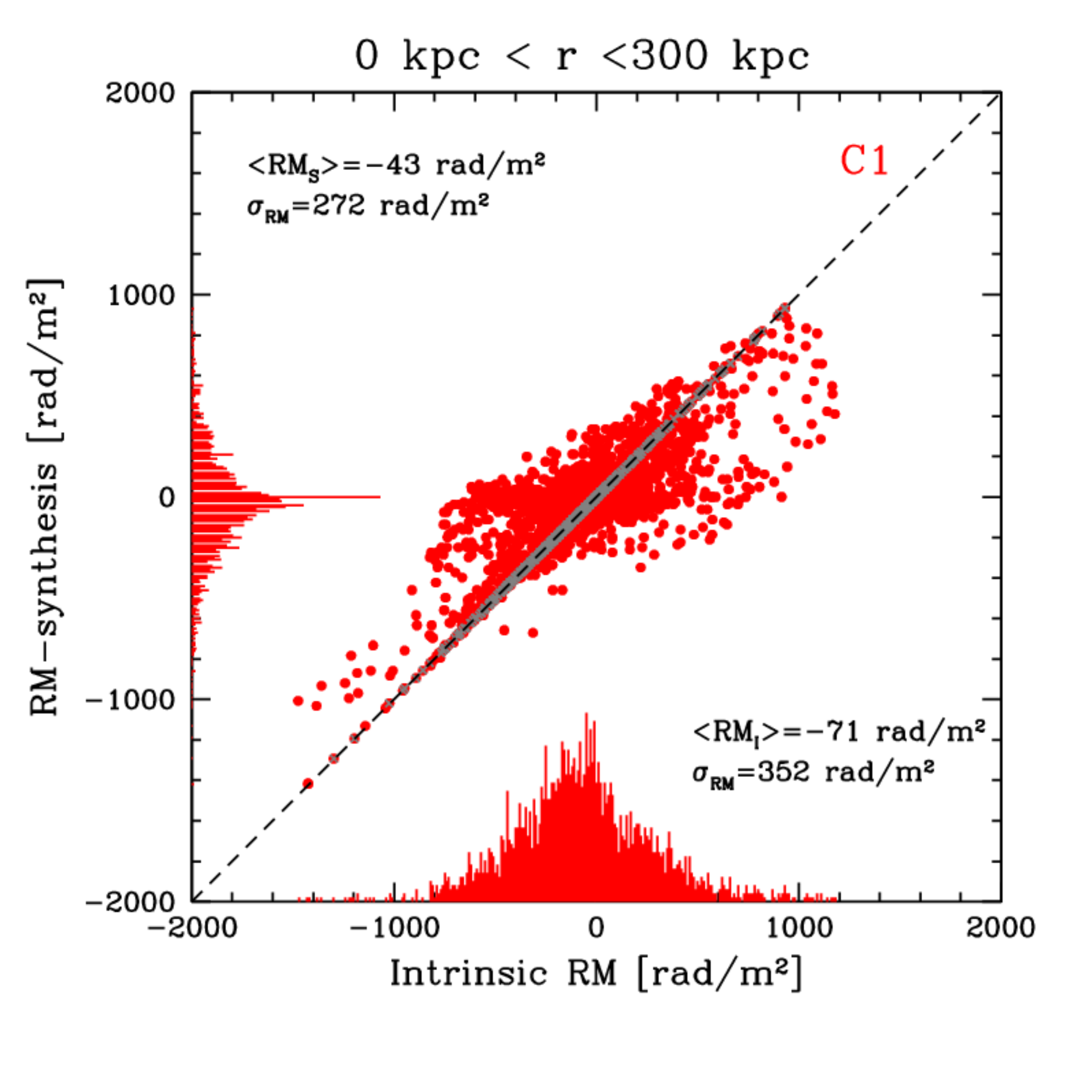}
    \caption{Point-to-point comparison between 0 and 300\,kpc from the cluster centre of the intrinsic $RM$ values (x-axis) and those from $RM$-synthesis (y-axis) for the simulated C1 cluster without radio halo. Distribution histograms, along with mean and rms, are also given. Grey crosses indicate the contribution of background sources.}
    \label{fig:histo}
\end{figure}
\subsection{The $\sigma_{RM}$ profile}
The $\sigma_{RM}$ profiles are computed from the Faraday depth images (bottom images of Figs. \ref{fig:rmsynth_img} and \ref{fig:rmsynth_img_noise}) in annuli of 6 pixel of width corresponding to a spatial width of $\sim$64\,kpc ($\sim30\arcsec$). The uncertainties associated to the measurements of the output Faraday depth images are calculated as follows \citep{sotomayor}:
\begin{equation}
    \Delta\phi(l)=\frac{1}{S/N}\frac{\sqrt{3}}{\lambda_2^2-\lambda_1^2},
    \label{eq:err}
\end{equation}
being $\lambda_2$ and $\lambda_1$ the maximum and minimum wavelengths of the bandwidth, respectively, and S/N the signal to noise ratio (here, equal to 1).
Therefore, the uncertainty on $\sigma_{RM}$ is:
\begin{equation}
    \Delta \sigma_{RM}= \sqrt{\sum_i \left ( \frac{\partial\sigma_{RM}}{\partial \phi_i(l)} \Delta \phi_i(l) \right )^2}= \frac{\Delta \phi(l)}{\sqrt{N}},
\end{equation}
being N the number of pixels considered to compute the corresponding $\sigma_{RM}$.\\
Figs. \ref{fig:rm_prof} and \ref{fig:rm_prof_noise} show the $\sigma_{RM}$ profile from respectively simulated and synthetic images, respectively. Top and bottom panels refer to the C1 and C2 simulated clusters. They are, from left to right, cluster without radio haloes, with radio haloes in equipartition, and simulated coupling between non-thermal and thermal particles. In each panel a solid magenta line represent the profile computed from the intrinsic $RM$ images (Fig. \ref{fig:rm}), and points and dotted lines refer to the profile computed from the output Faraday depth (bottom images of \ref{fig:rmsynth_img}).\\
Simulated data show some deviations with respect to the intrinsic $\sigma_{RM}$ values near to the cluster centre.
These central decrements are due to a different distribution of the intrinsic $RM$ values with respect to the $RM$-synthesis $RMs$, as shown in Fig. \ref{fig:histo} for the simulated C1 cluster without radio halo. In this plot the $RM$ values are compared point to point within a distance of 300\,kpc from the cluster centre. The $RM$ distributions are projected onto the axis, along with the mean and the rms of each distribution. The $RM$-synthesis histogram is more peaked near to zero than the intrinsic one, which is consistent with having lower values of $\sigma_{RM}$ at distances less than 300\,kpc. Clusters with radio haloes show similar histograms, included the C2 cluster hosting a radio halo for which the intrinsic and simulated $\sigma_{RM}$ are alike. For all the three cases under study, the two histograms start to be more consistent at distances larger than 300\,kpc, where the intrinsic $RM$ values are lower.
The source position with respect to the cluster centre is probably  responsible for the observed discrepancies. Gray crosses in Fig. \ref{fig:histo} indicate the dots that refer to background sources which cover the $\sim$30\% of the cluster area within a radius equal to 300\,kpc. In this plot the crosses overlap each other over the $RM$ value range creating a continuous grey region along the linear correlation drawn as a dashed black line. The perfect match between the intrinsic and $RM$-synthesis $RM$ values support the claim that the discrepancies are only due to the different paths crossed by the linearly polarized signals along the line-of-sight. \\

\begin{figure}
    \centering
    \includegraphics[width=0.47\textwidth]{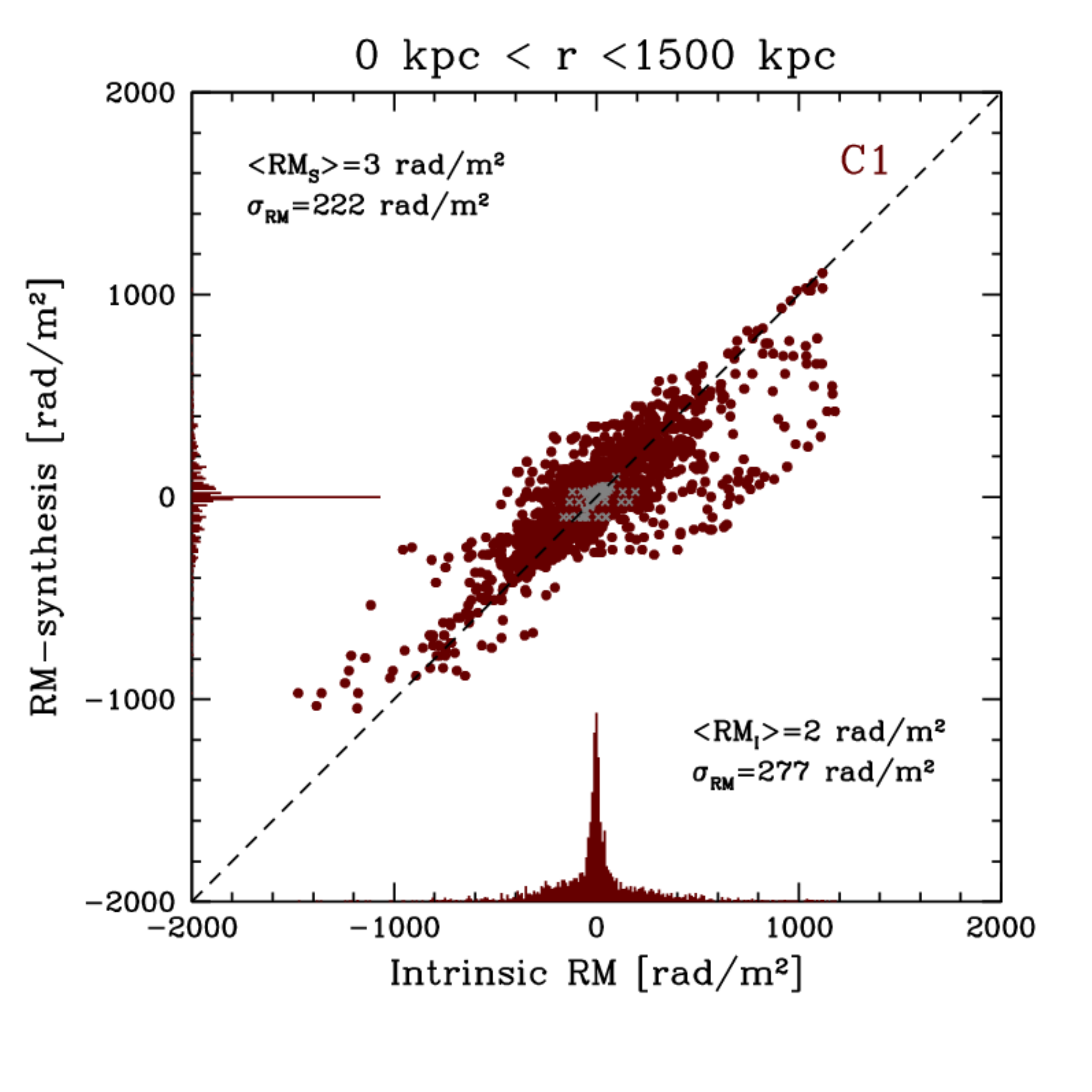}
    \caption{Point-to-point comparison between 0 and 1500\,kpc from the cluster centre of the intrinsic $RM$ values (x-axis) and those from $RM$-synthesis (y-axis) for the synthetic C1 cluster without radio halo. Distribution histograms, along with mean and rms, are also given. Grey crosses indicate the contribution of background sources.}
    \label{fig:histo2}
\end{figure}

Going from simulated to synthetic data, the $\sigma_{RM}$ profiles change dramatically for clusters without radio haloes compared to clusters with radio haloes.
In simulated data the entire cluster area is covered by background and cluster radio sources and, as shown in Fig. \ref{fig:histo}, the presence of the background sources guarantees the recovery of the cluster $RM$. When a thermal noise is added and the images are convolved at 10$\arcsec$ the background sources are too faint to survive in polarized images mainly because of the beam depolarization. Therefore, the measured $RM$ is mostly associated to the cluster sources as already mentioned in Section \ref{sect:p}. Indeed, within a distance of 300\,kpc from the C1 cluster centre there are not $RMs$ associated to background sources, and they cover $\sim$7\% of the cluster area and up to a radius of 1500\,kpc. Moreover, in this condition the background sources do not report the correct $RM$ values as shown in Fig. \ref{fig:histo2}. 
The under-sampling of the cluster area, mainly due to the low S/N for background sources, and the source position can justify the irregular behaviour of the synthetic $\sigma_{RM}$ profiles of clusters without radio haloes, where it is possible to observe both higher and lower values of $\sigma_{RM}$ with respect to the intrinsic one.\\
\begin{figure*}
    \centering
    \includegraphics[width=0.49\textwidth]{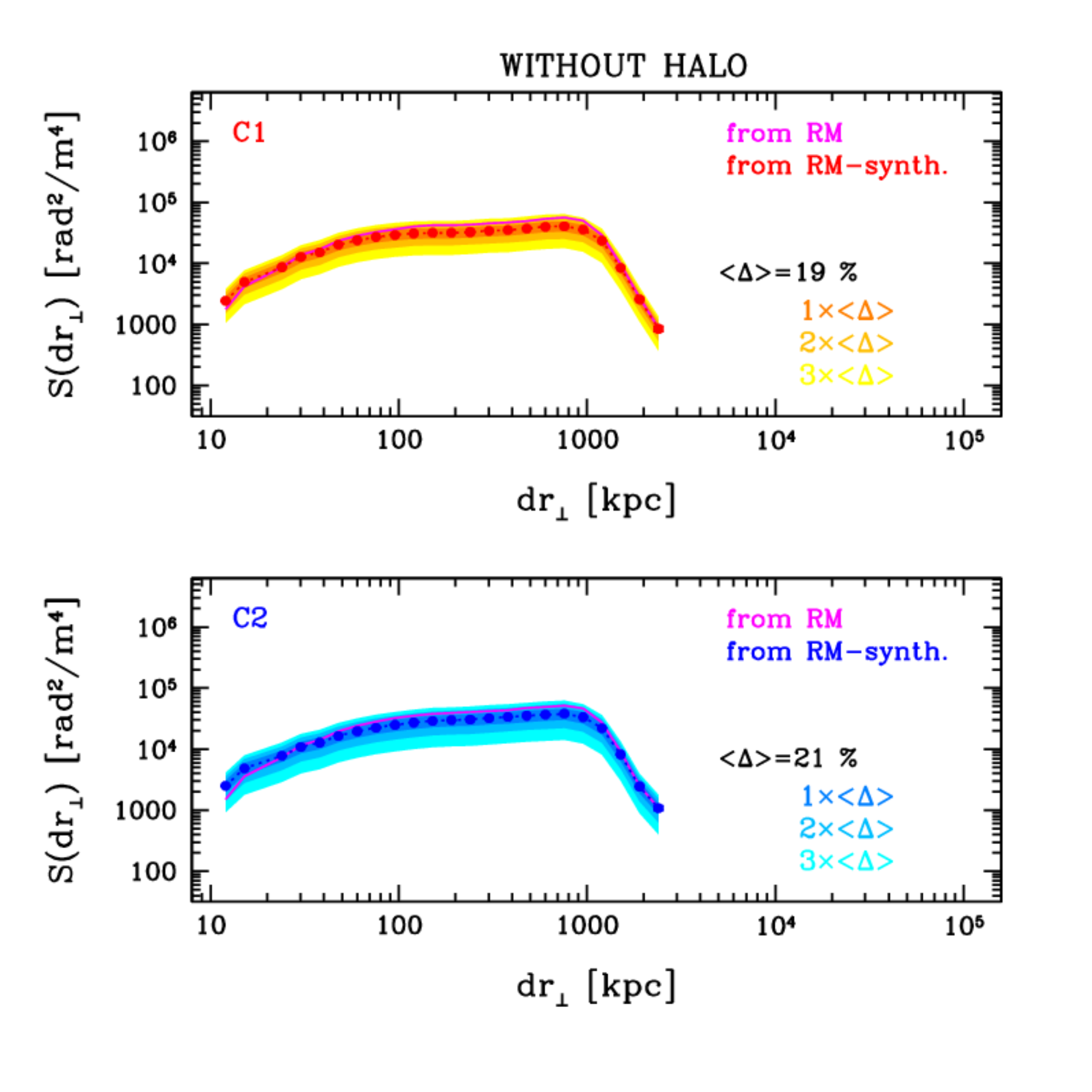}
    \includegraphics[width=0.49\textwidth]{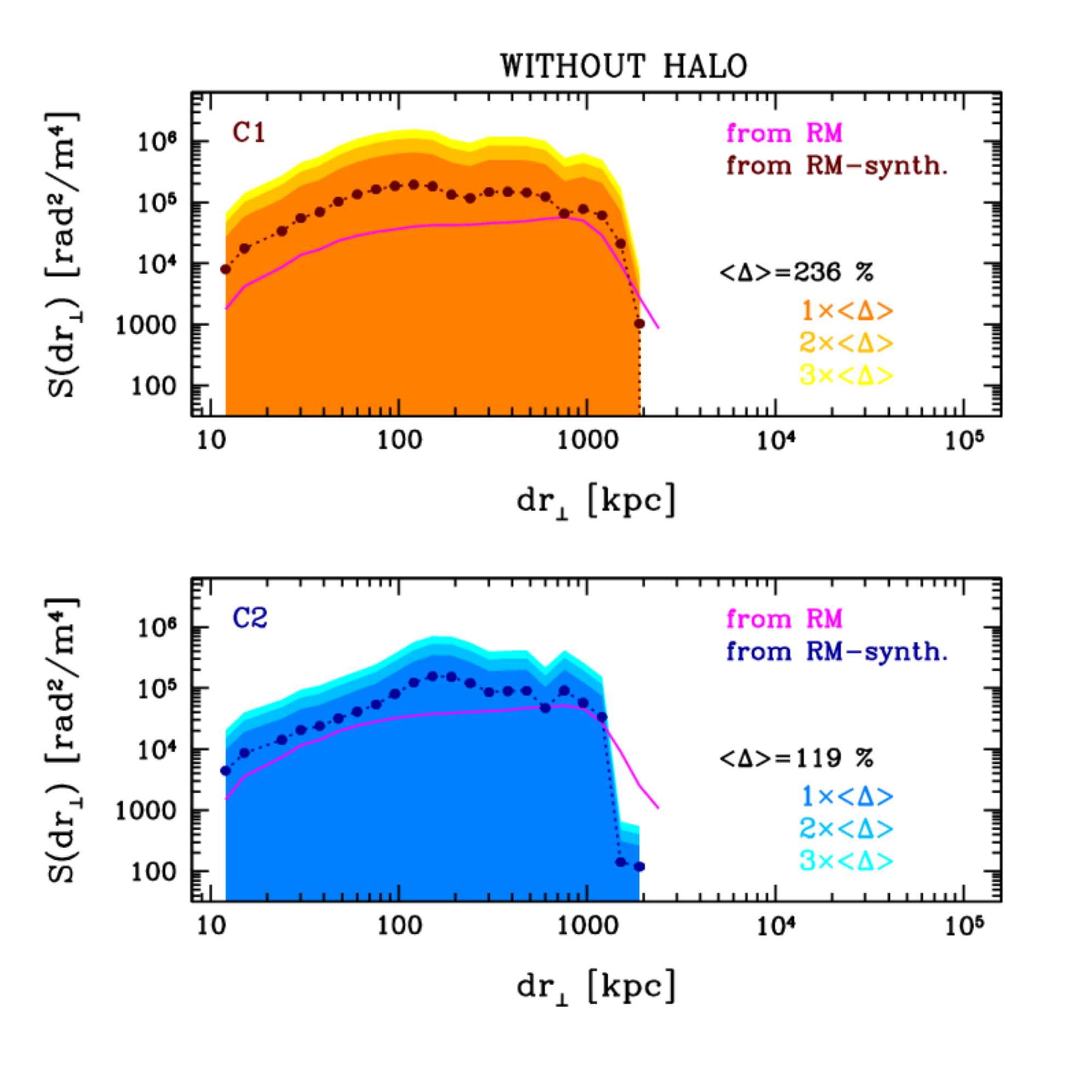}
   \caption{$RM$ structure function of clusters without radio haloes respectively for the C1 (top) and C2 clusters (bottom). On the left, the results obtained on simulated data, on the right, the results for synthetic data. In each panel, magenta solid lines represent the structure function computed from the intrinsic $RM$ images (Figs. \ref{fig:rm}, and \ref{fig:rmsynth_img_noise}) and points and dotted red lines refer to the structure function computed from the output Faraday depth (right images of Figs. \ref{fig:rmsynth_img} and \ref{fig:rmsynth_img_noise}). $\langle \Delta \rangle$ indicate the average difference in absolute value between the results of the $RM$-synthesis and of the intrinsic $RM$ (magenta line).}
    \label{fig:sdr_nh}
\end{figure*}
Even cluster hosting radio haloes show a synthetic profile different from the simulated one, more evident in the case of the C2 rather than the C1 cluster, and in particular $\sigma_{RM}$ presents lower values near to the centre. Here, the sampling problem is not observed, and so we believe that this effect is due both to the presence of a thermal noise, which as observed in the case of background sources can shift the peak in the Faraday dispersion function, and also to the convolution of the $Q$ and $U$ data cubes with a Gaussian function having FWHM=10$\arcsec$, which decreases the polarized intensity and therefore increases the shifts produced by the thermal noise. These effects clearly take place also in the case of clusters without radio haloes.\\
To quantify the precision with which data trace the intrinsic $RM$, it is useful to compute the average difference in percentage between the results of the $RM$-synthesis $\sigma_{RM}^{out}$ and the intrinsic $\sigma_{RM}^{in}$:
\begin{equation}
    \langle \Delta \rangle =\frac{1}{N}\sum_i^N \frac{|\sigma_{RM}^{out}-\sigma_{RM}^{in} |}{\sigma_{RM}^{in}},
\end{equation}
considering the N data points available in each annulus.
The $\langle \Delta \rangle$ values are shown on the right of the panels and shadow region are traced at 1, 2, and 3$\times\langle\Delta\rangle$.\\
Simulated data show mean differences between $\sim$8-15\% with respect to the $\sigma_{RM}^{in}$ values where the smallest values are in correspondence of clusters without radio haloes. For synthetic data, the differences are of $\sim$18-51\% with respect to the input values, where the largest values occurs for clusters without radio haloes.\\
From these results, it is clear that the $\sigma_{RM}$ profile retrieved with the procedure proposed here is a good diagnostic tool for cluster magnetic fields measurements within $\sim$20-30\% when a diffuse source is hosted by the cluster independently from the scenario which triggers its emission and within $\sim$50\% for clusters without radio haloes.
\begin{figure*}
    \centering
    \includegraphics[width=0.49\textwidth]{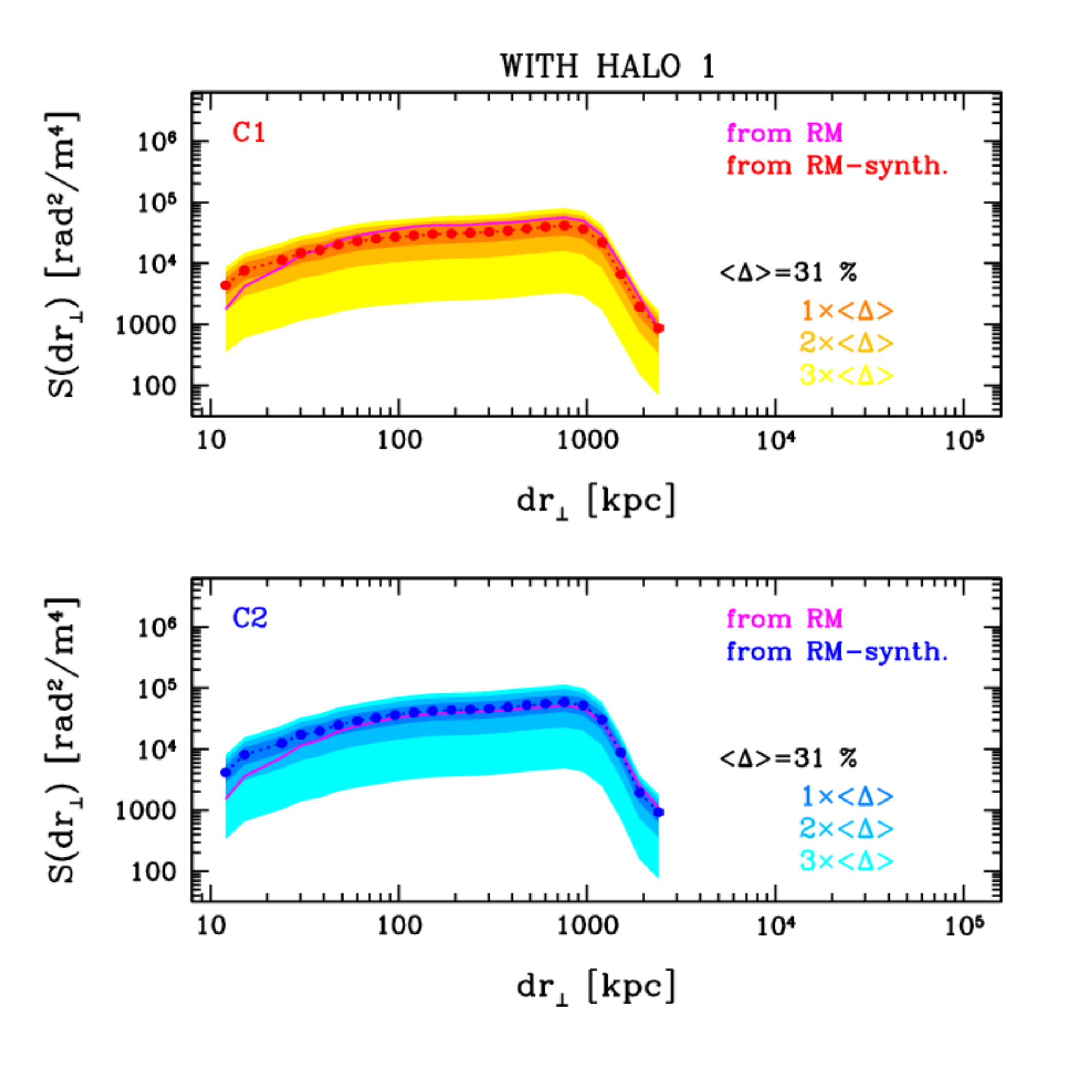}
    \includegraphics[width=0.49\textwidth]{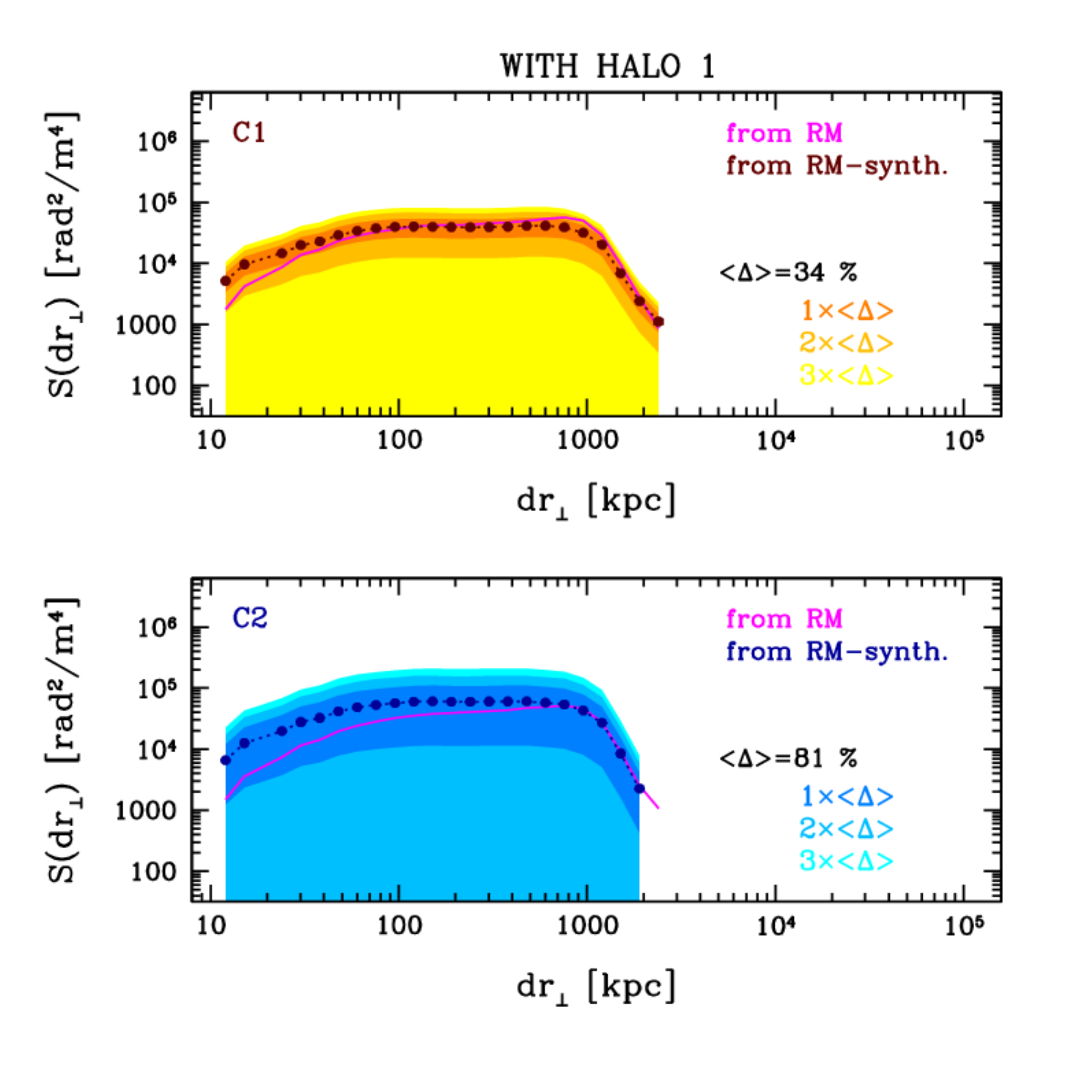}
    \caption{$RM$ structure function of clusters with radio haloes in equipartition ({\it HALO 1}). On the left, the results obtained on simulated data are shown while on the right, the results for synthetic data respectively for the C1 (top) and C2 clusters (bottom). In each panel, magenta solid lines represent the structure function computed from the intrinsic $RM$ images (Fig. \ref{fig:rm}), and \ref{fig:rmsynth_img_noise}) and points and dotted red lines refer to the structure function computed from the output Faraday depth (central images of \ref{fig:rmsynth_img} and \ref{fig:rmsynth_img_noise}). $\langle \Delta \rangle$ indicate the average difference in absolute value between the results of the $RM$-synthesis and of the intrinsic $RM$ (magenta line).}
    \label{fig:sdr_h}
\end{figure*}

\subsection{The $RM$ Structure function}
The $RM$ structure function is defined as:
\begin{equation}
    S(dr)=\langle [RM(r)-RM(r+dr)]^2 \rangle_{(x,y)},
\end{equation}
where $r$ is the distance in kpc and $S(dr)$ in rad$^2$/m$^4$ and $\langle \dots \rangle_{(x,y)}$ means that the average is taken over all the positions $(x,y)$ in the $RM$ image.\\
The structure function is computed in a circle of 1\,Mpc of radius, centred on the cluster centres, with a logarithmic sampling from $\sim$10\,kpc up to the maximum scale of $\sim$2\,Mpc. By propagating the $RM$ errors, the uncertainty associated with the structure function measurements {\rm becomes}:
\begin{equation}
    \Delta S(dr)= \frac{2 \sqrt{2S(dr)}}{\sqrt{N}} \Delta \phi(l),
\end{equation}
where $\Delta\phi(l)$ has been defined in Eq. \ref{eq:err} and N is the number of measurements for each value of $S(dr)$.\\
Fig. \ref{fig:sdr_nh} shows the resulting structure functions for the C1 (top) and C2 (bottom) clusters without the radio halo signal. The left panels refers to the simulated data, while on the right the results obtained considering the synthetic data are plotted. In each panel, magenta solid lines represent the structure function computed from the intrinsic $RM$ images (Fig. \ref{fig:rm}), and points and dotted red lines refer to the structure function computed from the output Faraday depths (right images of Figs. \ref{fig:rmsynth_img} and \ref{fig:rmsynth_img_noise}). The error bars are comparable with the size of the dots. In the same way, Fig. \ref{fig:sdr_h} shows the resulting structure functions for clusters with radio haloes simulated assuming the equipartition condition.
The results concerning the radio haloes obtained coupling between the thermal and non-thermal particle energy densities are similar to what obtained for radio haloes in equipartition.\\
The $RM$ structure functions present an increase up to scales $\sim$100-200\,kpc, followed by a flat behaviour with a very smooth increase and a decrease at $\sim$1\,Mpc, due to the poor sampling on large scales. \\
Going from simulated to synthetic data, it is possible to observe an irregular trend of the $RM$ structure function traced by clusters without radio haloes. 
As for the $\sigma_{RM}$ profile, this could be related to the beam depolarization and to the presence of the thermal noise. As a consequence synthetic data sample a smaller portion of the cluster area with respect to simulated data.
Indeed, when the diffuse sources are considered, it is noticeable that the synthetic data follow the intrinsic $RM$ structure function much more faithfully.\\
As for the $\sigma_{RM}$ profile, mean differences $\langle\Delta\rangle$ between the data and the intrinsic $RM$ structure function has been evaluated:they are shown in the bottom right corner of each panel. For clusters without radio haloes, the resulting values go from a minimum of $\sim$20\,\% in the case of the simulated data up to a maximum of $\sim$236\% in the case of the synthetic C1 cluster. In the case of clusters with radio haloes, the mean discrepancies with respect to the intrinsic value are of $\sim$30\% for simulated data, while for synthetic data are of the order of $\sim$35-80\%.\\  
Shadow regions have been drawn for reference at 1, 2, and 3$\times\langle\Delta\rangle$. We note that the intrinsic profiles lies within the 2$\times\langle\Delta\rangle$ region of synthetic data with haloes while the profile for clusters without radio haloes is within the 3$\times\langle\Delta\rangle$ region up to a distance of $\sim$1\,Mpc. The $\langle\Delta\rangle$ values clearly show that a good and homogeneous sampling of the $RM$ is necessary to trace the intrinsic $RM$ structure function. 

\section{Conclusions}
In this work, the $RM$-synthesis technique has been applied on synthetic SKA1-MID radio images of a pair of galaxy clusters at 1.4\,GHz with a resolution of 10$\arcsec$ and a thermal noise of 0.1\,$\muup$Jy/beam. \\
The ICM properties have been produced from a cosmological MHD simulations and analysed. Using the \textsc{faraday} software package the clusters' total intensity and the linearly polarized emission were modelled. This emission is related to a population of cluster, background and foreground radio galaxies and star forming galaxies. Two models have been considered to simulate diffuse synchrotron sources at the centre of the clusters: one assumes equipartition between the relativistic particles energy density and the magnetic field energy density on one side, and the other forces the relativistic particles energy density to be a fraction of 0.3\% of the thermal particle energy density.\\
$RM$-synthesis has been applied in order to determine the polarized intensity of the sources and the cluster $RM$. The results were compared to the input information for both the clusters and for the three simulated cases (clusters without radio haloes, clusters with radio haloes in equipartition, clusters with radio haloes simulated coupling between thermal and non-thermal particles), considering either the polarized intensity or the $\sigma_{RM}$ profile or the $RM$ structure function.\\
The results can be summarise as follows:
\begin{enumerate}
    \item the two models assumed for the radio haloes produce emissions both in total intensity and in polarization with different extension and morphology. Indeed, radio haloes in equipartition show a more pronounced filamentary emission which extend at large distances from the cluster centre with a discontinuous arrangement of bright and faint filaments while radio haloes simulated assuming coupling between non-thermal and thermal plasma energy density show a smoother emission rapidly decreasing going to the outskirts.
    These differences are useful to assess the nature of relativistic particles in radio haloes, which will become possible with the advent of next generation radio telescope such as the SKA.
    \item Even if their characteristics are different, the two kind of cluster radio haloes do not show significant discrepancies in $RM$ images, $\sigma_{RM}$ profiles and $RM$ structure functions.
    \item The comparison between the linearly polarized signal at $RM$=0 and those resulting from the application of the $RM$-synthesis demonstrate the efficiency of this technique in recovering polarization, especially at the centre of the clusters where the high values of $RM$ can significantly depolarize signals with a percentage of $\sim$40-60\%.
    \item Synthetic data trace the instrinsic $\sigma_{RM}$ profiles, within a percentage of $\sim$20-30\% for clusters hosting radio haloes and of $\sim$50\% for clusters without radio haloes. A good sampling of the cluster area is of fundamental importance to constrain the cluster $\sigma_{RM}$ and the presence of diffuse polarized radio sources can play a key role in the determination of the cluster $RM$ properties.
    \item Similar conclusion can be drawn for the $RM$ structure function. Synthetic data of clusters without radio haloes reconstruct an irregular trend of this function and present average differences between 120 and 240\% with respect to the intrinsic values. On the other hand, the presence of a radio halo, independently from the model assumed to reproduce its emission, can guarantee a better estimation of the intrinsic $RM$ structure function with discrepancies of the order of 30-80\%. 
\end{enumerate}

It is worth recalling that the results obtained in this work are model dependent.
The discrete radio sources have been modelled with very recent radio luminosity functions and on the basis of what is known about their polarization properties \citep[see][for more details]{loi19}. Despite the high expectations of the $RM$ measurements from polarized extra-galactic sources with the SKA, this study shows that with a resolution of 10$\arcsec$ and a thermal noise in $Q$ and $U$ data of 0.1$\,\muup$Jy/beam, the majority of background sources do not emerge from the noise and cannot therefore give us a $RM$ value.\\
With the SKA1-MID several surveys have been proposed to be conducted at 0.5$\arcsec$ and 2$\arcsec$ of resolution. 
Polarized images at such resolutions are less affected by beam depolarization and could drive to different results, with intrinsic $\sigma_{RM}$ profiles and $RM$ structure function traced in a better way by synthetic data. To investigate this point MHD cubes of higher spatial resolution are needed and this will be the subject of a future work.\\
To conclude, it is important to notice the key role that diffuse radio sources can play in the determination of the $RM$ and therefore of the intracluster magnetic field properties.

\section*{Acknowledgements}
We gratefully acknowledge the anonymous referee for the useful comments and suggestions. We also thanks E. Bonnasieux which carefully read the manuscript helping us to improve it.
FL and AB acknowledge financial support from the Italian Minister for Research and Education (MIUR), project FARE SMS, code R16RMPN87T. AB acknowledges financial support from the ERC-Stg DRANOEL, no 714245. IP acknowledges funding from the INAF PRIN-SKA 2017 project 1.05.01.88.04 (FORECaST). IP, FL and MM acknowledge support from INAF under the MAIN STREAM PRIN project "SAUROS".
The trg computer cluster was funded by the Autonomous Region of Sardinia (RAS) using resources from the Regional Law 7 August 2007 n. 7 (year 2015) "Highly qualified human capital", in the context of the research project CRP 18 "General relativity tests with the Sardinia Radio Telescope" (P.I. of the project: Dr. Marta Burgay).







\appendix
\section{Single scale magnetic field models compared with data}
\begin{figure}
    \centering
    \includegraphics[width=0.47\textwidth]{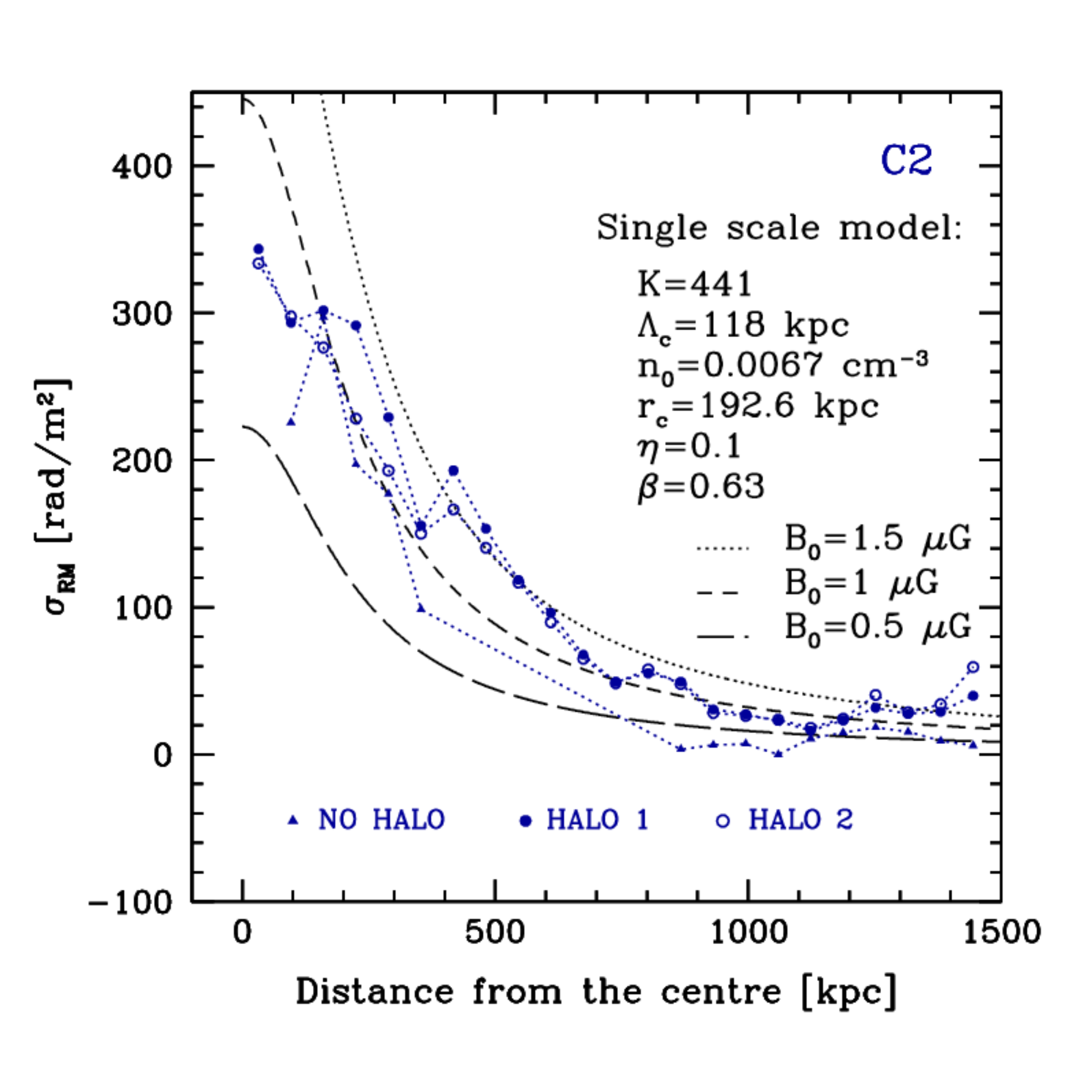}
    \caption{Comparison of the synthetic $\sigma_{RM}$ profile of the C2 cluster without radio halo (triangles), with radio halo in equipartition (full circles) and coupled with the thermal plasma (open circles) with the single scale magnetic field model assuming $B_0=0.5, 1, 1.5\,\muup G$.}
    \label{fig:felten}
\end{figure}

The $\sigma_{RM}$ profile and the $RM$ structure function are useful to determine the power spectrum of the intracluster magnetic field and therefore its strength and structure. To do so, it is necessary to apply advanced computational techniques which, starting from a modelling of the thermal plasma and of the intracluster magnetic field, produce synthetic images of the $RM$ and by comparing these images with observations can determine the magnetic fields characteristics. This interpretative part is beyond the scope of this work and it would require a very long computational time \citep[see e.g.][]{govoni17}. However, it is possible to give a rough estimate of the intracluster magnetic field strength in a simple way. \\
A magnetic field tangled on a single scale $\Lambda_C$ randomly oriented from cell to cell results in a Gaussian-like $RM$ distribution with zero mean and dispersion $\sigma_{RM}$ \citep{law,felten}. Assuming a $\beta$-model for the thermal density (see Eq. \ref{eq:bmodel}), and the magnetic field strength radial profile to be a function of the thermal gas density:
\begin{equation}
    B = B_0 \left ( \frac{n}{n_0} \right )^{\eta},
\end{equation}
where $\eta$ and $B_0$ are the exponential factor and the central value of the magnetic field, the $\sigma_{RM}$ as a function of the projected distance r is \citep{dolag}:
\begin{equation}
   \sigma_{RM}(r)=K B_0 \Lambda_c^{0.5} n_0 r_c^{\frac{1}{2}} \frac{1}{(1+\frac{r^2}{r_c^2})^{\frac{6\beta(1+\eta)-1}{4}}} \sqrt{ \frac{\Gamma[3\beta(1+\eta)-\frac{1}{2}]}{\Gamma[3\beta(1+\eta)]}}.
   \label{eq:fel}
\end{equation}
The factor K is a constant which depends on the integral path and in particular K=624 for background sources and K=441 for sources lying in the middle of the cluster.\\
Fig. \ref{fig:felten} shows the comparison between the $\sigma_{RM}$ synthetic data of the C2 cluster and the single-scale model.
To trace the profile of Eq. \ref{eq:fel} the values reported in Section 2 have been used, namely $\Lambda_c=118$\,kpc, $n_0=0.0067\,cm^{-3}$, $r_c=192.6$\,kpc, $\beta=0.63$. The fit of the $\log(B/B_0)-\log(n/n_0)$ profiles between 100 and 500\,kpc yields a power law with index $\eta=0.1$. Assuming K=441 three profiles are plotted considering $B_0=0.5, 1, 1.5\,\muup G$. 
In all the three cases, the data are better described with the single scale model with $B_0=1\,\muup$G and indeed the C2 intracluster magnetic field has a central strength of $\sim$0.8-0.9\,$\muup$G (see Fig. \ref{fig:ps}).


\bsp	
\label{lastpage}
\end{document}